\documentclass[8pt]{article}
\usepackage{BarryStyleFile}

\usepackage{bbold}
\usepackage{indentfirst}
\usepackage{graphicx,wrapfig,float,slashed}
\usepackage{amsmath,amssymb,epsfig,graphicx,xcolor}
\usepackage{epstopdf}
\usepackage{ragged2e}
\usepackage{relsize}
\usepackage{tikz}
\usepackage{enumerate}
\usepackage{cancel}
\usepackage{lipsum}
\usepackage{epstopdf}
\usepackage{xcolor}
\usepackage{tabu}
\usepackage[toc]{appendix}
\usepackage[normalem]{ulem}
\usepackage[font=small,skip=1pt]{caption}
\allowdisplaybreaks
\usepackage{pifont}
\usepackage{subfig}
\usepackage{amsmath}

\newcommand{\bea}{\begin{eqnarray}}
\newcommand{\eea}{\end{eqnarray}}
\newcommand{\beq}{\begin{equation}}
\newcommand{\eeq}{\end{equation}}
\newcommand{\ba}{\begin{align}}
\newcommand{\ea}{\end{align}}
\newcommand{\bpm}{\begin{pmatrix}}
\newcommand{\epm}{\end{pmatrix}}
\newcommand{\non}{\nonumber}
\newcommand{\trm}[1]{\textrm{#1}}

\newcommand{\gp}{g_{\phi e}}
\newcommand{\pp}{p^{\prime}}
\newcommand{\cd}{\cdot}
\newcommand{\ep}{\epsilon}
\newcommand{\varphip}{\varphi^{\prime}}

\title{\huge ALP production through non-linear Compton scattering in intense fields}
\author{Barry M. Dillon and Ben King}

\affiliation{\centerline{\textit{\footnotesize{Centre for Mathematical Sciences, Plymouth University, Plymouth PL4 8AA, United Kingdom}}}}
\emailAdd{b.dillon@plymouth.ac.uk}
\emailAdd{b.king@plymouth.ac.uk}

\abstract{We derive production yields for massive pseudo-scalar and scalar axion-like-particles (ALPs), through non-linear Compton scattering of an electron in the background of low- and high-intensity electromagnetic fields. In particular, we focus on electromagnetic fields from Gaussian plane wave laser pulses.
A detailed study of the angular distributions and effects of the scalar and pseudo-scalar masses is presented.
It is shown that ultra-relativistic seed electrons can be used to produce scalars and pseudo-scalars with masses up to the order of the electron mass. 
We briefly discuss future applications of this work towards lab-based searches for light beyond-the-Standard-Model particles.
}

\begin{document}

\maketitle
\flushbottom

\section{Introduction}  \label{intro}

\noindent The existence of light exotic pseudo-scalar or scalar particles is highly motivated by some of the most widely studied extensions of the Standard Model (SM).
For example, the strong-CP problem of QCD can be elegantly solved through the Peccei-Quinn mechanism \cite{PQ1} which predicts the existence of a new spin-0, CP-odd particle called the axion\footnote{see \cite{AxionTheoryReview} for review articles.}. 
Many other extensions of the SM also contain spontaneously broken $U(1)$ global symmetries and light spin-0 Goldstone bosons as a result.
We broadly refer to these light spin-0 particles as axion-like-particles (ALPs) irrespective of their CP properties.
The interactions of these particles are described by the Lagrangian densities
\begin{align} \label{ALPints}
\mathcal{L}^{\text{int}}_-=&~-m^2_{\phi}\phi^2-\frac{g_{\phi\gamma\gamma}}{4}\phi F^{\mu\nu}\widetilde{F}_{\mu\nu}-g_{\phi e}\phi\bar{\Psi}\gamma^5\Psi		\non\\
\mathcal{L}^{\text{int}}_+=&~-m^2_{\phi}\phi^2-\frac{g_{\phi\gamma\gamma}}{4}\phi F^{\mu\nu}F_{\mu\nu}-g_{\phi e}\phi\bar{\Psi}\Psi
\end{align}
where the $\pm$ refers to the even/odd CP property of the ALP, $\phi$ represents the ALP field, $F^{\mu\nu}$ is the electromagnetic field strength tensor, and $\widetilde{F}^{\mu\nu}=\tfrac{1}{2}\ep^{\mu\nu\eta\delta}F_{\eta\delta}$ is the dual field strength tensor.
A vast amount of work has been done studying the experimental signatures\footnote{for a recent summary of the experimental status we refer the reader to \cite{GrahamReview}, and for a recent review including proposals for new experiments we refer the reader to \cite{Redondo2018}.} of these fields in low-energy lab-based experiments (light-shining-through-wall (LSW) experiments) \cite{LabExp}, solar experiments \cite{SolarExp}, dark matter and stellar evolution experiments \cite{DMExp}, beam dumps \cite{BDumpExp}, rare meson decays \cite{MesonExp}, and in high energy collider experiments \cite{ColliderExp}.
In the LSW experiments, and in most other ALP searches, the aim is to produce and detect ALPs through ALP-photon conversion mediated by the $g_{\phi\gamma\gamma}$ coupling (for a review on theoretical aspects of LSW experiments see \cite{LSWtheoryReview}).

Unique opportunities for observing beyond-the-SM (BSM) processes are also found through the study of quantum electrodynamics (QED) in high-intensity fields.
The calculation of scattering matrix elements in high intensity fields cannot be performed using a perturbative expansion in the coupling and instead one must use non-perturbative solutions of the Dirac equation to describe the interaction between the electromagnetic field and the electron.
The most popular of these solutions, of which very few are known, is the Volkov solution for an electron in a plane wave background \cite{Volkov}.
For reviews on these methods we refer the reader to \cite{Ritus1985,SFQEDreview} and we list recent developments in the field in \cite{SFQEDrecent}.
Given the increasing availability of high-intensity lasers from recent and upcoming experiments \cite{LaserExp,SarriExperiments} the study of QED in intense fields to observe both SM and BSM phenomena is of crucial importance. So far, using high-intensity lasers to probe BSM phenomena has mainly be studied theoretically, and then through the coupling to ALPs \cite{ALPslaser} and mini-charged paticles \cite{MCPslaser} to photons. In this paper we study interactions involving electrons and ALPs in intense electromagnetic fields, focusing on the production of ALPs via Compton scattering from electrons in intense laser pulses.
We will make some reasonable assumptions on the parameters in the calculation; the first being that the laser photons have an energy of $\kappa^0=1.55$eV (corresponding to a wavelength of $800$nm), and the second being that the electrons can have momenta of up to a few GeV in optical set-ups. (In colliding $47$ GeV electrons with a $\trm{ps}$ optical laser pulse, the SLAC E144 experiment \cite{slac} is an example of combining particle accelerator and laser pulse technology.) We will assume that a bunch of electrons interact incohorently with the external field, and for that reason restrict ourselves to processes involving single electron seeds. (Optical set-ups typically deal with bunches of the order of $\sim10^8$ electrons \cite{SarriExperiments}.)

We begin the paper with a study of the Compton production of ALPs in a head-on collision between the seed electron and a low-intensity laser pulse. Due to the large number of photons, even in this low-intensity example, the laser pulse can still be  treated as a classical background field and we expand the electron wavefunction perturbatively in a small intensity parameter, $\xi \ll 1$. 
This dimensionless intensity parameter represents the work done by the external field over the Compton wavelength of an electron, in units of the external field photon energy and so in some way quantifies the number of photons interacting at a time with an electron. $\xi$ will be defined quantitatively at the beginning of the next section.
We assume the laser background to have a Gaussian pulse shape, however we also assume that the pulse duration is much larger than the photon wavelength, allowing us to approximate the electron as being in a monochromatic background.
After obtaining analytical expressions for the production yield of the ALPs in both a linearly and circularly polarised laser pulse we study the total yield and angular distribution of the emitted ALPs for various ALP masses.
We then move on to the study of electron-ALP interactions in high-intensity fields, i.e. $\xi\gg1$.
Using the Volkov solution for the electron wave-function we take the limit of a constant-crossed background field and calculate the production yield of the ALPs via non-linear Compton scattering.  A similar calculation for the emission of a massless pseudoscalar was performed in \cite{Borisov}, where bounds on the ALP properties were derived using astrophysical constraints.
Employing the Locally Constant Field Approximation (LCFA), see for example \cite{lcfa,DiPiazza1}, we use this result to approximate the production yield of ALPs when the background electromagnetic field has a non-trivial profile - such as a Gaussian or that of a focussed laser pulse.
Using these solutions, we present a detailed analysis of the energy and angular distribution of the production yield for the ALPs.
We study the effects of having a non-zero mass term for the ALPs and perform a comparison between the properties of scalar and pseudo-scalar production.
Finally we conclude and discuss how this work can be used in studies of lab-based searches for light ALPs which probe the  $g_{\phi e}$ coupling.

\section{ALP production in a low-intensity laser pulse}  \label{perturbative}

\noindent In this section we study the Compton production of a scalar or pseudo-scalar from an electron in a low intensity external electromagnetic field.
The external electromagnetic field is parametrised by
\beq \label{gaugea}
A^{\mu}(x)=\frac{a^{\mu}(x)}{e}=\frac{m_e\xi}{e}\ep^{\mu}f(x)
\eeq
where $\xi=eA^0/m_e$ is the dimensionless intensity parameter, and $f(x)$ is a pulse shape which describes the spatial dependence of the vector potential (the parameter $\xi$ can be defined in a gauge- and Lorentz-invariant manner using the stress-energy tensor, see \cite{IldertonHeinzl2009}). The low-intensity regime is then defined by the intensity parameter satisfying $\xi \ll 1$.
We choose a linearly polarised external field in the $\ep^{\mu}=(0,1,0,0)^\mu$ direction, and label the perpendicular polarisation as $\tilde{\ep}^{\mu}=(0,0,1,0)^\mu$.
The photon momenta are described by $\kappa^\mu=\kappa^0(1,0,0,1)^\mu$, and we define a dimensionless phase $\varphi=\kappa\cd x$ which we use to parametrise the position of the electron wavefunction with respect to the external field.
In the low intensity regime, the effects of this field on the electron wavefunction can be treated perturbatively, i.e. we expand the wavefunction to first order in $a^{\mu}$ where the zeroth order part contains the free electron wavefunction and the first order part contains the interaction.
Thus we have
\begin{align}
\psi_{p,r}=&\psi_{p,r}^{(0)}+\psi_{p,r}^{(1)} \non\\
\psi_{p,r}^{(0)}=&\frac{e^{-ip\cd x}}{\sqrt{2Vp^0}}u_r(p),~~\psi_{p,r}^{(1)}=-i\int d^4y ~G(x\!-\!y)\cancel{a}(y)\psi_{p,r}^{(0)}(y)
\end{align}
where the $G(x\!-\!y)=\langle0|T\psi(x)\psi(y)|0\rangle=\int\tfrac{d^4p}{(2\pi)^4}\tfrac{i(\cancel{p}+m)}{p^{2}-m_e^{2} + i\varepsilon}e^{-ip\cdot(x-y)}$ is the fermionic propagator.
This method approximates that only one photon is absorbed by the electron prior to the emission of the pseudo-scalar or scalar particle.
The matrix element for the process $e^-\rightarrow\phi+e^-$ can be written as
\beq
S_{fi}=-i\gp\int d^4x ~\phi_k\bar{\psi}_{p^\prime,r^{\prime}}\gamma^5\psi_{p,r}
\eeq
where $p$, $p^{\prime}$, and $k$ are the momenta of the incoming and outgoing electrons and the outgoing scalar or pseudo-scalar, respectively.
The $r$ and $r^{\prime}$ label the spinor indices of the incoming and outgoing electrons. Then $S_{fi} = S^A_{fi}+ S^B_{fi}$ where
\begin{align}
S^A_{fi}=&-i\gp\int d^4x~\phi_k\bar{\psi}^{(0)}_{\pp,r^{\prime}}\gamma^5\psi^{(1)}_{p,r},~~~\text{and}	\non\\
S^B_{fi}=&-i\gp\int d^4x~\phi_k\bar{\psi}^{(1)}_{\pp,r^{\prime}}\gamma^5\psi^{(0)}_{p,r}.
\end{align}
The matrix element for scalar production is obtained by replacing $\gamma^5$ by the spinor identity matrix as indicated by the structure of the interaction in Eq. \ref{ALPints}.
The outgoing wavefunction of the scalar or pseudo-scalar field is
\beq
\phi_k=\frac{e^{-ik\cd x}}{\sqrt{2Vk^0}}.
\eeq
The probability can then be written in the form
\begin{align} \label{prob1}
P=&\frac{1}{V^2}\int\frac{d^3p^{\prime}d^3k}{(2\pi)^6}\sum_{\text{spin}}\trm{tr}|S^A_{fi}+S^B_{fi}|^2\non\\
P=&\frac{(m_e\xi)^2\gp^2}{2^4(\kappa^0)^2(2\pi)^3}\int \frac{d^2k^{\perp}dk^-}{p^-p^{\prime-}k^-}\theta(p^-\!-\!k^-)\theta(k^-)\Bigg|\tilde{f}\left(\frac{k^++p^{\prime+}-p^+}{2\kappa^0}\right)\Bigg|^2\mathcal{T}_{\pm}
\end{align}
where $\mathcal{T}_{\pm}$ contains the traces over the spinor indices for either the scalar or the pseudo-scalar interaction.
Note here that we have used the lightfront coordinates for the particle momenta, a description of which can be found in Appendix \ref{appA}.
The spatial dependence of the external field now enters through the Fourier transform of the pulse shape, $\tilde{f}$, and its argument follows from momentum conservation imposed on the momentum of the recoiling electron, 
\beq 
p^{\prime-}=p^--k^-,~~~p^{\prime\perp}=p^{\perp}-k^{\perp},~~~p^{+}=\frac{(p^{\perp})^2+m_e^2}{p^-},
\eeq 
where the last expression is simply the on-shell condition.

\subsection{The pulse shape and the monochromatic limit}

\noindent To obtain the expression in Eq. \ref{prob1} we began by Fourier transforming the profile $f(x)$ as
\beq
a_{\mu}(x)=\ep_{\mu}m_e\xi\int\frac{dr}{2\pi}\tilde{f}(r)e^{-ir\kappa\cdot x},
\eeq
where $\kappa^{\mu}=\kappa^0(1,0,0,1)^{\mu}$ describes a plane wave trajectory for the photon with $\kappa^0$ being the photon energy.
We suppose that the pulse shape for $f(x)$ is Gaussian with respect to the phase $\varphi$, i.e.
\beq\label{LaserPulse}
f(\varphi)=\frac{1}{2}\left(e^{-\left(\frac{\varphi}{\Phi}\right)^2+i\varphi}+e^{-\left(\frac{\varphi}{\Phi}\right)^2-i\varphi}\right)
\eeq
where $\Phi=\kappa\cd\tau$ is a pulse duration with $\tau^\mu=\tau^0(1,0,0,0)^\mu$ in the lab frame, and the terms linear in $\varphi$ in the exponent describe the oscillations of the plane wave with frequency $\kappa^0$.
From this we can calculate
\beq
\tilde{f}(r)=\int d\varphi~f(\varphi)e^{ir\varphi}=\frac{\sqrt{\pi}}{2}\Phi\left(e^{-(r+1)^2\frac{\Phi^2}{4}}+e^{-(r-1)^2\frac{\Phi^2}{4}}\right)
\eeq
and insert it into Eq. \ref{prob1}.
This results in a complicated expression which can be simplified by assuming that  $\Phi\gg1$.
Using this we arrive at the monochromatic (or long-pulse) limit
\beq
\frac{\tilde{f}(r)^2}{\Phi}\simeq \left(\frac{\pi}{2}\right)^{3/2}\left[\delta(r+1)+\delta(r-1)\right].
\eeq
Taking
\beq\label{rdef}
r=\frac{k^++p^{\prime+}-p^+}{2\kappa^0}
\eeq
in Eq. \ref{prob1}, the long pulse limit implies that the incoming electron absorbs or emits one photon of fixed energy to or from the external field.
After these manipulations the probability can be written as
\begin{align}
P=&~\frac{(m_e\xi)^2\gp^2}{2^5(\kappa^0)^2(2\pi)^3}\left(\frac{\pi}{2}\right)^{3/2}\Phi\int \frac{d^2k^{\perp}dk^-}{p^-p^{\prime-}k^-}\theta(p^-\!-\!k^-)\theta(k^-)[\delta(r\!-\!1)+\delta(r\!+\!1)]~\mathcal{T}
\end{align}
where one of the integrals over $d^2k^\perp dk^-$ will be used to enforce the delta function condition.

\subsection{Simplifying the expressions for the yields}

\noindent 
The function $\mathcal{T}_{\pm}$ can be written as
\beq
\mathcal{T}_{\pm}=\frac{\alpha_{\pm}}{(2\pp\cd k\!+\!m_{\phi}^2)^2}+\frac{\beta_{\pm}}{(2\pp\cd k\!+\!m_{\phi}^2)(m_{\phi}^2\!-\!2k\cd p)}+\frac{\gamma_{\pm}}{(m_{\phi}^2\!-\!2k\cd p)^2}
\eeq
where $\pm$ corresponds to scalar and pseudo-scalar, respectively.
Before writing these factors it is useful to note that momentum conservation implies
\begin{align} \label{momCon}
p\cd\pp=&~m_e^2-p\cd k+\kappa\cd p~r		\non\\
=&~m_e^2-\tfrac{m_{\phi}^2}{2}+\kappa\cd k~r		\non\\
p\cd k=&~ \tfrac{m_{\phi}^2}{2}+(\kappa\cd p-\kappa\cd k)r	\non\\
\pp\cd k=&~ -\tfrac{m_{\phi}^2}{2} +\kappa\cd p~r
\end{align}
where we recall that $r$ is the variable from the Fourier transformation of $f(\varphi)$ defined in Eq.~\ref{rdef}.
Using these relations we can write the factors in the trace as
\begin{align}
\alpha_-=&~8\left(r^2\kappa\cd p(\kappa\cd p-\kappa\cd k)+2r(\ep\cd p)(\ep\cd k)\kappa\cd p-(\ep\cd p)^2m_{\phi}^2\right)	\non\\
\beta_-=&~16\left(r^2\kappa\cd p(\kappa\cd p-\kappa\cd k)-r(\ep\cd k)^2\kappa\cd p-(\ep\cd p)^2m_{\phi}^2+(\ep\cd p)(\ep\cd k)(m_{\phi}^2+2r\kappa\cd p-r\kappa\cd k)\right)\non\\
\gamma_-=&~8r^2\kappa\cd p(\kappa\cd p-\kappa\cd k)+16r(\ep\cd k)(\ep\cd p-\ep\cd k)(\kappa\cd p-\kappa\cd k)-8m_{\phi}^2(\ep\cd p-\ep\cd k)^2	\non\\
\alpha_+=&~\alpha_-	 +32(\ep\cd p)^2m_e^2\non\\
\beta_+=&	~\beta_- +64(\ep\cd p)(\ep\cd p-\ep\cd k)m_e^2	\non\\
\gamma_+=&~\gamma_- +32(\ep\cd p-\ep\cd k)^2m_e^2.
\end{align}
We see that when the electrons collide head-on with the laser pulse i.e. $(\ep\cd p)=(\tilde{\ep}\cd p)=0$, the expressions simplify greatly and the scalar and pseudo-scalar kinematics differ only in the $\gamma_\pm$ terms.
From here onwards, we focus solely on the case in which $(\ep\cd p)=(\tilde{\ep}\cd p)=0$.
The next step is to perform the $k$ integrals.
The first point to note is that the integrations force $r=1$, because the negative $r\!=\!-\!1$ solution is kinematically forbidden.
The second point to note is that we can do the integrals either in $d^2k^\perp dk^-$ or in $d^2k^\perp dk^3$.
And lastly the third point to note is that the integrand is independent of $\tilde{\ep}\cd k$, apart from through $r$ which is a function of $(\ep\cd k)^2+(\tilde{\ep}\cd k)^2$.
We define $\ep\cd k=m_e\rho\cos\phi$ and $\tilde{\ep}\cd k=m_e\rho\sin\phi$ with $0\leq\phi\leq2\pi$ and $\rho\geq0$, such that the delta function simplifies to
\begin{align}
\delta(r-1)=2\frac{\kappa^0 k^-(p^-\!-\!k^-)}{m_e^2p^-}\frac{1}{\sqrt{g}}\delta\left(\rho-\sqrt{g}\right)
\end{align}
with
\beq
g=2\frac{\kappa^0 k^-(p^-\!-\!k^-)}{m_e^2p^-}-\left(\frac{k^-}{p^-}\right)^2-\frac{p^-\!-\!k^-}{p^-}\frac{m_{\phi}^2}{m_e^2}.
\eeq
We can now write the probability as
\begin{align} \label{Ppert2}
P=&~\frac{(m_e\xi)^2\gp^2}{2^4\kappa^0(2\pi)^3}\left(\frac{\pi}{2}\right)^{3/2}\Phi\int \frac{d\rho~ d\phi~ dk^-}{(p^-)^2}\theta(p^-\!-\!k^-)\theta(k^-)\delta(\rho-\sqrt{g})	\frac{\rho}{\sqrt{g}}\mathcal{T_\pm}.
\end{align}
To simplify the expressions we define $k^-=v p^-$, $m_{\phi}=\delta m_e$, and $\kappa\cd p=\eta_p m_e^2$.  Performing the $\rho$ and $\phi$ integrals we have
\begin{align}
P_{\pm}=&\frac{1}{\eta_p}\frac{\xi^2 \gp^2}{2^4(2\pi)^3}\left(\frac{\pi}{2}\right)^{3/2}\Phi\int_{0}^{v_{\text{max}}}dv~ \mathcal{T}_{\pm}	\non\\
\mathcal{T}_-=&~4\pi\left(\frac{2v^2}{1-v}\right)-\delta^2\frac{4\pi}{\eta_p}\frac{2v}{1-v}+\delta^2\frac{4\pi}{\eta_p^2}\left(\frac{\delta^2}{1-v}+\frac{v^2}{(1-v)^2}\right)	\non\\
\mathcal{T}_+=&~4\pi\left(\frac{2v^2}{1-v}\right)-(\delta^2-4)\frac{4\pi}{\eta_p}\frac{2v}{1-v}+(\delta^2-4)\frac{4\pi}{\eta_p^2}\left(\frac{\delta^2}{1-v}+\frac{v^2}{(1-v)^2}\right).
\end{align}
Note that we require $g>0$ to obtain real solutions, putting a limit on $v$ which can be written as
\beq
2\eta_p-\frac{\delta^2}{v}-\frac{v}{1-v}\geq0,~~~0\leq v\leq1.
\eeq
This ensures $v<1$ and thus the expressions for $\mathcal{T}_{\pm}$ never become singular.
In the case where we use a circularly polarised external field we find that the result is the exact same.
This is only true however in the limit where $\ep\cd p=\tilde{\ep}\cd p=0$.

\subsection{Angular distributions and effects of a non-zero ALP mass} \label{DistPert}
\begin{figure}[h!!]
\centering
\includegraphics[width=9cm]{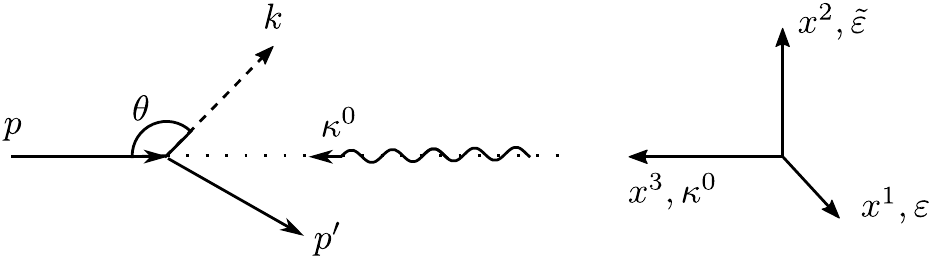}
\caption{A schematic diagram of the interaction taking place between the laser background and electron, leading to a scattered electron and ALP.\label{Polar}}
\end{figure}

\begin{figure}[H]
\centering
  \includegraphics[width=150mm]{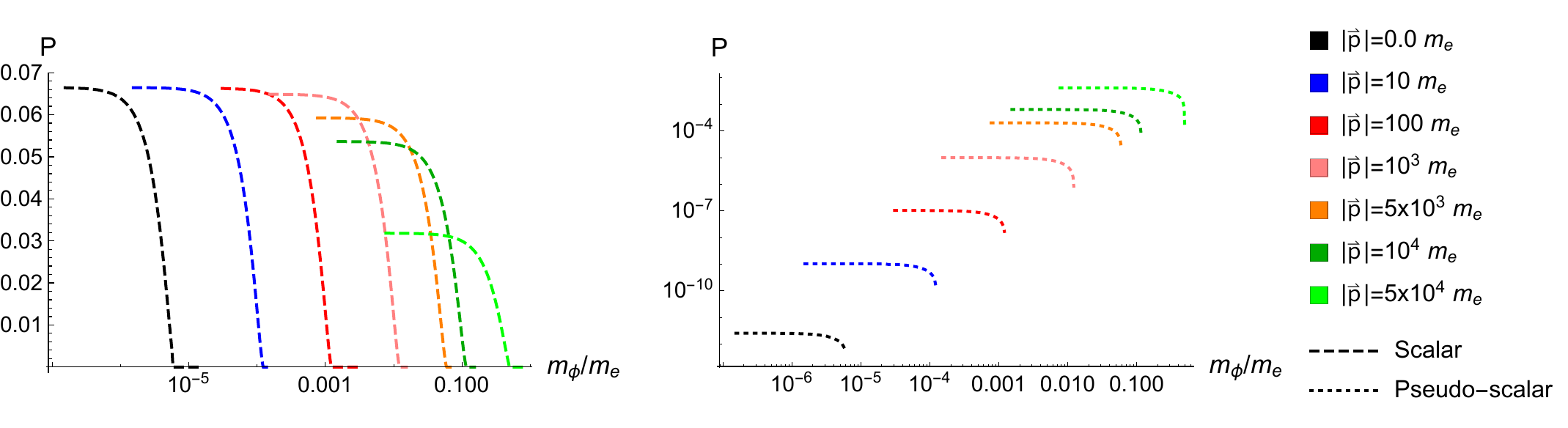}
\caption{The total yield is plotted against the mass of the emitted ALP for various seed electron energies.  \label{TYmassPert}}
\end{figure}

\begin{figure}[H]
\centering
  \includegraphics[width=80mm]{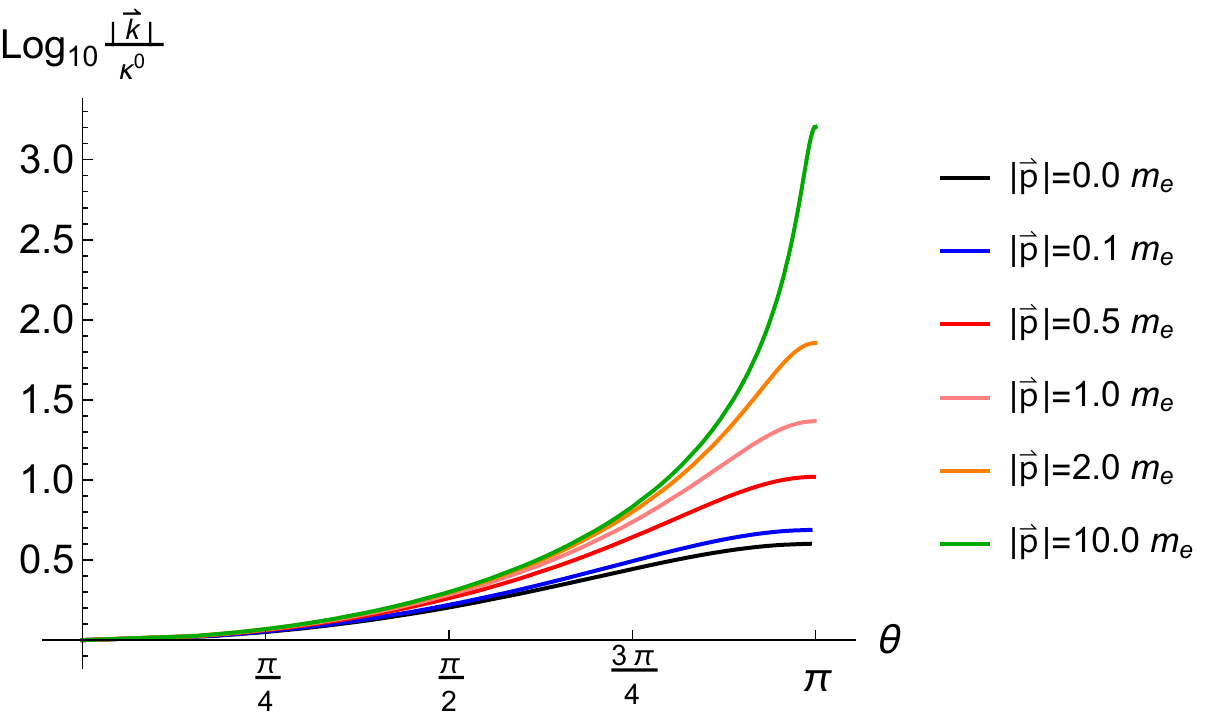}
\caption{Relationship between the energy of the emitted ALP and the polar angle at which it is emitted in lab frame.  These plots are for $m_{\phi}=0$.  \label{angpert}}
\end{figure}

\begin{figure}[H]
\centering
  \includegraphics[width=80mm]{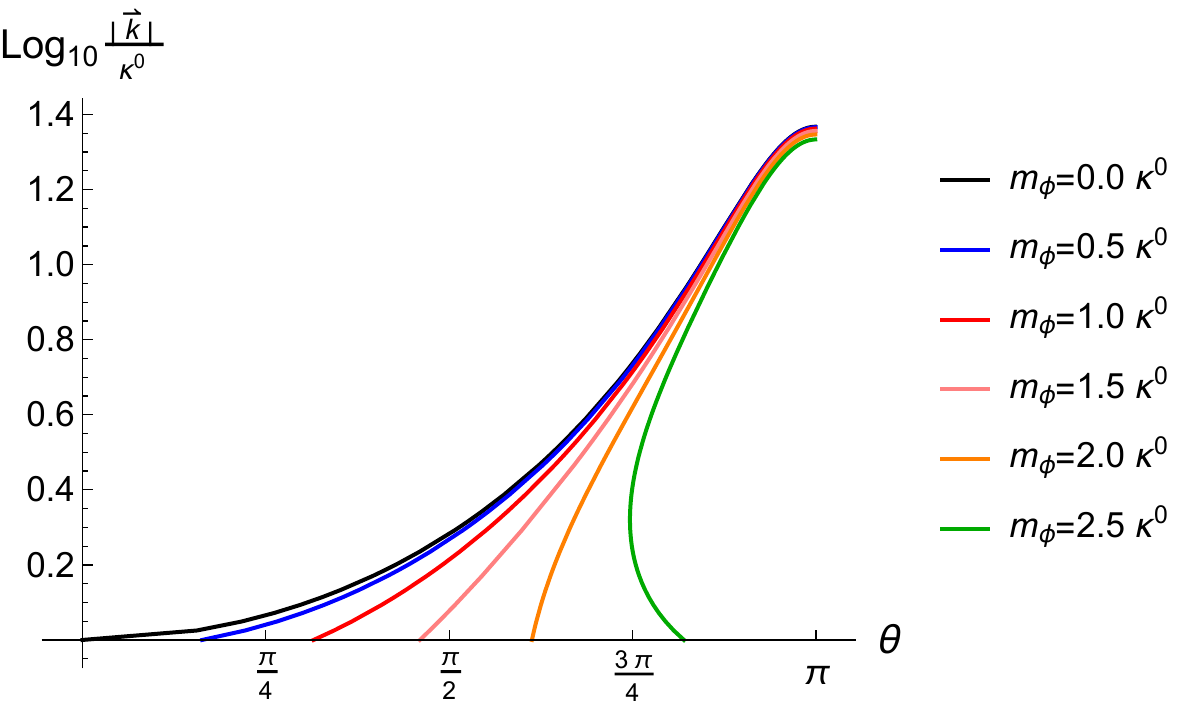}
\caption{Relationship between the energy of the emitted ALP and the polar angle at which it is emitted in lab frame for $|\vec{p}|=m_e$ and various ALP masses.\label{angpertmass}}
\end{figure}

\noindent  As mentioned in the introduction, $\kappa^0=1.55$eV and $|p|\lesssim\mathcal{O}(10)$GeV, from which it follows that $\eta_p\ll1$. 
(This can be seen from $\eta_p=\tfrac{\kappa^0}{m_e}(\sqrt{1+(|\vec{p}|/m_e)^2}+|\vec{p}|/m_e)$).
In Figure \ref{TYmassPert} we plot the total yield for pseudo-scalar and scalar production as a function of the ALP mass for various seed electron momenta in the MeV range\footnote{For $\xi\ll1$ the parameters $\xi,~\Phi$, and $g_{\phi}$ all enter only as pre-factors and thus we set them to $1$ in the plots in this section.}.
We refer to the quantity $P$ as the production yield as it represents the number of ALPs expected to be emitted while the electron is in the external field.
When dealing with the external field $P$ can obviously not represent a probability since if $\Phi$ is large enough $P$ can be larger than $1$.
In a realistic experimental set-up the interaction would take place between a laser pulse and a bunch of approximately $10^8$ electrons.
We can see from Figure~\ref{TYmassPert} that the range of ALP masses probed by this interaction depends entirely on the energy of the initial seed electron, with the photon energy, $\kappa^0$, being fixed to $1.55$ eV.
We can see that the production yield cuts off quite sharply as $\delta$ reaches a critical value that depends on the seed electron energy.
With larger $|\vec{p}|$ values the cut-off increases approximately linearly, and $|\vec{p}|\sim\mathcal{O}$(GeV) allows the interaction to probe $m_{\phi}\sim2\times 10^4~\kappa^0\sim\mathcal{O}(m_e)$ masses.
It is useful to look at the scattering in lab-frame polar coordinates (see Appendix \ref{angular} for details on the co-ordinate transformation).  
The exact expressions of the yield in these co-ordinates are lengthy and we refrain from presenting them here, although the reader can easily deduce them from the information already provided.
In the monochromatic limit there is a direct relationship between the ALPs' energy and the polar angle at which they are emitted.
This is irrespective of the scalar or pseudo-scalar nature of the particle and we have plotted the relationship for various seed electron energies in Figure \ref{angpert}.
From this plot we can see that for $|\vec{k}|\ll\kappa^0$ the polar angle of emission is close to zero, which corresponds to emission parallel to the momentum of the incoming laser photon.
For larger ALP momentum the polar angle shifts closer to $\pi$, which corresponds to emission parallel to the momentum of the incoming electron.
As the seed electron energy increases the polar distribution becomes more sharply localised towards the polar angle of the incoming electron\footnote{See Figure~\ref{Polar} for schematic diagram showing the set-up we consider.}.
Permitting the ALP to have a non-zero mass drastically alters the properties of the emission, as can be seen in Figure \ref{angpertmass}.
In particular, a non-zero mass alters the polar angle at which the ALP is emitted and restricts it to lay closer to the polar angle along which the initial seed electron travels.
Note that with $\kappa^0=1.55$ eV we only reach $\eta_p=1$ when $|\vec{p}|\sim40$ GeV.
We have not plotted the $|\vec{k}|$ distribution of the emitted ALPs for the simple reason that the distributions are approximately constant over the ranges depicted in Figure \ref{angpert}, a feature which persists even for seed electron energies of $\mathcal{O}$(GeV).

\section{ALP production in a constant-crossed field}  \label{strong}

\noindent  If the intensity of the external field is such that 
 $\xi\gtrsim\mathcal{O}(1)$, the approximation that the electron absorbs only one photon before emitting the pseudo-scalar or scalar particle breaks down. All orders of photon exchange with the electron must be included for a calculation to be consistent.
In this section we calculate the production yield for pseudo-scalar and scalar particles via non-linear Compton scattering in a constant-crossed external field.
This result is particularly important as in the rest frame of an ultra-relativistic particle all electromagnetic fields resemble a constant-crossed field \cite{Ritus1985}.

We will use the notation and structures of the previous section, and introduce new ideas along the way.
The relevant non-linear parameter that we use in the study of high-intensity fields is $\chi_q=\xi\eta_q$, sometimes referred to as the \emph{quantum nonlinearity parameter}, which is equal to the work done by the external field over a Compton wavelength in units of the electron rest energy. Assuming $\kappa^0=1.55$~e and $\xi\sim\mathcal{O}(100)$ we have $\chi_q\gg\eta_q$ and to generate an $\mathcal{O}(1)$ $\chi_p$ for the seed electron we only require $|\vec{p}|\sim\tfrac{100}{\xi}$ GeV. In comparison, for non-relativistic electrons with $|\vec{p}|\lesssim\mathcal{O}$(MeV) and the same laser parameters, $\chi\sim6\times10^{-3}$.

To calculate the scattering matrix when $\xi\sim\mathcal{O}(1)$ or larger we must use the non-perturbative Volkov solution of the Dirac equation for an electron in a plane-wave background,
\beq
\psi_{p,r}=\left(1+\frac{\cancel{\kappa}\cancel{a}}{2\kappa\cd p}\right)e^{-ip\cdot x+iS^{\prime}_p}\frac{u_r(p)}{\sqrt{2p^0V}}.
\eeq
The dynamics of the background field is described by the $\cancel{\kappa}\cancel{a}$ and the $S^{\prime}_p$ terms, which are non-linear in the vector potential:
\beq
S^{\prime}_p=-\int_{-\infty}^{\varphi}dz\left(\frac{p\cd  a(z)}{\kappa\cd  p}-\frac{a^2(z)}{2\kappa\cd  p}\right),
\eeq
and $a^{\mu}(z)$ is the external plane-wave EM field defined as
\beq
a^{\mu}(z)=m_e\xi\ep^{\mu}g(\kappa\cd z),\quad\ep^2=-1,
\eeq
with $\kappa^{\mu}=\kappa^0(1,0,0,1)^{\mu}$ being the photon momentum vector. Despite using a plane-wave solution of the Dirac equation the constant field limit can be taken in integrated expressions through $\kappa^0\rightarrow0$.
We will see that the result for the total yield in the constant field is independent of $\kappa^0$ and the limit is trivial.
We will start with the calculation for pseudo-scalar production and then present the result for scalar production, this allows us to describe the calculation in more detail.
The matrix element is written in a similar way:
\beq
S_{fi}=-i\frac{\gp}{\sqrt{8V^3k^0p^0p^{\prime0}}}\int d^4x~e^{i(k+\pp-p)\cdot x}~\Gamma_-(\varphi),\quad\varphi=\kappa\cd  x=\kappa^0x^-
\eeq
The $\Gamma_-(\varphi)$ function arises from the Volkov solution and contains all the spinor and external field dependence
\beq
\Gamma_-(\varphi)=\bar{u}_{r^{\prime}}(\pp)\left(1+\frac{\cancel{a}\cancel{\kappa}}{2\kappa\cd  p^{\prime}}\right)\gamma_5\left(1+\frac{\cancel{\kappa}\cancel{a}}{2\kappa\cd  p}\right)u_r(p)e^{-iS^{\prime}_{\pp}+iS^{\prime}_p}.
\eeq
Performing similar steps to the previous section and transforming to light-front coordinates we find that the total probability is given by
\begin{align} \label{P}
P_-=&V^2\int\frac{d^3\pp}{(2\pi)^3}\int\frac{d^3k}{(2\pi)^3}\theta\left(p^{\prime 0}\right)\theta\left(k^0\right)\frac{1}{2}\sum_{\text{spins}}\trm{tr}|S_{fi}|^2	\non\\
=&\frac{\gp^2}{2^4(2\pi)^3(\kappa^0)^2}\int\frac{d^2k^\perp dk^-}{k^-p^-p^{\prime-}}\sum_{\text{spins}}\trm{tr}~\widetilde{\Gamma}_-^{\dagger}(r)\widetilde{\Gamma}_-(r)
\end{align}
with
\begin{align}
\sum_{\text{spins}}\trm{tr}~\widetilde{\Gamma}_-^{\dagger}(r)\widetilde{\Gamma}_-(r)=\int &d\varphi d\varphip ~T(p,k,\pp,\varphi,\varphip )	\non\\
&e^{i\int_{\varphi^{\prime}}^{\varphi}dz\left[r+\alpha m_e\xi g(z)+\beta(m_e\xi)^2g^2(z)\right]}
\end{align}
originating from the Fourier transform of $\Gamma^\dagger_-\Gamma_-$.
The function $T(p,k,p^{\prime},\varphi,\varphi^\prime)$ contains the spinor traces that arise after taking the spin sum.
The $\alpha$ and $\beta$ functions from the above expression are given by
\beq
\alpha=\left(\frac{\ep\cd \pp}{\kappa\cd \pp}-\frac{\ep\cd p}{\kappa\cd p}\right),~~~\beta=\frac{1}{2}\left(\frac{1}{\kappa\cd \pp}-\frac{1}{\kappa\cd p}\right).
\eeq
The trace element of the spin sum is contained within the $T(p,k,\pp,\varphi,\varphip )$ function, which can be simplified using the momentum conservation relations in Eq. \ref{momCon} to find
\begin{align}
\sum_{\text{spins}}\trm{tr}\,\tilde{\Gamma}_-^{\dagger}(r)\tilde{\Gamma}_-(r)=\int &d\varphi~ d\varphip ~\Big[ 2m_\phi^2-2\kappa\cd k\Big(2r+(m_e\xi)\alpha[g(\varphi)+g(\varphi^{\prime})]+2\beta(m_e\xi)^2g(\varphi)g(\varphi^{\prime})\Big)\Big]	\non\\
&e^{i\int_{\varphi^{\prime}}^{\varphi}dz\left[r+\alpha m_e\xi g(z)+\beta(m_e\xi)^2  g^2(z)\right]}.
\end{align}

\noindent The spatial integrals can be computed exactly in the constant field limit, corresponding to $g(x)=x$. 
To perform the $\varphi$ and $\varphip$ integrals a change of variables is useful, and we choose
\beq
\varphi_+=\frac{1}{2}(\varphi+\varphip),~~~\varphi_-=(\varphi-\varphip).
\eeq
The integrals can be performed exactly using the integral identities in appendix~\ref{appB} and the total probability can be written as 
\begin{align} \label{P1}
P_-=\frac{2^{2/3}g_{\phi e}^2}{2^4 \pi}\int\frac{d^2k^{\perp}d\chi_k}{m_e^2}\Bigg[&\frac{1}{\chi_k^2}\left(\frac{\chi_k}{\chi_p(\chi_p\!-\!\chi_k)}\right)^{1/3}\left(\frac{\Delta}{m_e^2}+\frac{\chi_k^2}{\chi_p(\chi_p\!-\!\chi_k)}\right)\text{Ai}^2\left(\frac{\Sigma}{2^{2/3}}\right) \non\\
&+\frac{2^{2/3}}{\chi_k}\left(\frac{\chi_k}{\chi_p(\chi_p\!-\!\chi_k)}\right)^{2/3}\text{Ai}^{\prime2}\left(\frac{\Sigma}{2^{2/3}}\right)\Bigg]
\end{align}
where
\begin{align} \label{S1}
\Delta=&\frac{(\chi_k\tilde{\ep}\cd p - \chi_p\tilde{\ep}\cd k)^2}{\chi_p(\chi_p-\chi_k)}	\non\\
\Sigma=&\frac{1}{\chi_k}\left(\frac{\chi_p(\chi_p-\chi_k)}{\chi_k}\right)^{1/3}\Big[\frac{\Delta}{m_e^2}+\delta^2+\frac{\chi_k^2}{\chi_p(\chi_p-\chi_k)}\Big]
\end{align}
where we recall $\delta=m_\phi/m_e$. The above result is independent of $\ep\cd k$.
This is precisely because we have chosen the constant-crossed field background, which we will discuss in more detail in the next section.
Using the identities listed in Appendix \ref{appB} the $\tilde{\ep}\cd k$ integral can be performed exactly and we find
\begin{align} \label{P2}
P_-=-\frac{g_{\phi e}^2}{8\pi}\int_{-\infty}^{\infty}\frac{d(\ep\cd k)d\chi_k}{m_e}\Bigg[  \frac{1}{\chi_p}\left(\frac{\chi_k}{\chi_p(\chi_p\!-\!\chi_k)}\right)^{1/3} \text{Ai}^{\prime}(\Omega)+\frac{1}{2} \frac{\delta^2}{\chi_p\chi_k}\text{Ai}_1(\Omega)\Bigg]
\end{align}
where
\beq \label{S2}
\Omega=\left(\frac{\chi_k}{\chi_p(\chi_p\!-\!\chi_k)}\right)^{2/3}+\frac{\delta^2}{\chi_k}\left(\frac{\chi_p(\chi_p\!-\!\chi_k)}{\chi_k}\right)^{1/3}
\eeq
and $\text{Ai}_1(x)=\int_0^{\infty}\text{Ai}(t+x)dt$.
\newline

The calculation for the production of a scalar proceeds analogously, apart from the $\gamma^5$ operator is replaced with the spinor identity matrix in the interaction between the $\phi$ and fermion fields.
The fermion trace for the scalar field production is equal to that in the pseudo-scalar trace plus a factor of $8m_e^2$, and the terms in the exponent of the integrand in the spin sum remain unchanged.
Integrating the $(\varphi_+,\varphi_-)$ variables we find that the total probability can be written as
\begin{align} 
P_+=-\frac{2^{2/3}g_{\phi e}^2}{2^4 \pi}\int\frac{d^2k^\perp d\chi_k}{m_e^2}\Bigg[&\frac{1}{\chi_k^2}\left(\frac{\chi_k}{\chi_p(\chi_p\!-\!\chi_k)}\right)^{1/3}\left(4+\frac{\Delta}{m_e^2}+\frac{\chi_k^2}{\chi_p(\chi_p\!-\!\chi_k)}\right)\text{Ai}^2\left(\frac{\Sigma}{2^{2/3}}\right) \non\\
&+\frac{2^{2/3}}{\chi_k}\left(\frac{\chi_k}{\chi_p(\chi_p\!-\!\chi_k)}\right)^{2/3}\text{Ai}^{\prime2}\left(\frac{\Sigma}{2^{2/3}}\right)\Bigg]
\end{align}
where the form of $\Delta$ and $\Sigma$ in the $\chi_q$ notation can be found in Eq. \ref{S1}.
Integrating over $d(\tilde{\ep}\cd k)$ we then find,
\begin{align} \label{S2}
P_+=-\frac{g_{\phi e}^2}{8\pi}\int_{-\infty}^{\infty}\frac{d(\ep\cd k)d\chi_k}{m_e}\Bigg[  \frac{1}{\chi_p}\left(\frac{\chi_k}{\chi_p(\chi_p\!-\!\chi_k)}\right)^{1/3} \text{Ai}^{\prime}(\Omega)- \frac{2}{\chi_p\chi_k}\left(1-\frac{1}{4}\delta^2\right)\text{Ai}_1(\Omega)\Bigg].
\end{align}
It is important to notice here that the argument of the Airy function is the same as in the pseudo-scalar case, which follows from the kinematics of the collision.

\section{ALP production in high-intensity backgrounds}  \label{lcfa}
\noindent In the constant-crossed field calculation there appears a divergent integral in $\ep\cd k$. However, this can be reinterpreted in the following way. The integral over the electron's phase co-ordinate $\varphi$, performed at the amplitude level, leads to the Airy functions at the level of the probability. The contribution from the Airy functions occurs mainly when they have an argument of the order of unity or less. This corresponds to a finite region of the electron's trajectory in $\varphi$. This so-called ``coherence interval'' \cite{Ritus1985} becomes ever smaller as $\xi$ increases. In the limit $\xi \to \infty$, the relevant part of the electron trajectory corresponds to the stationary phase:
\beq \label{psistar}
\varphi_*=\frac{1}{m_e\xi}\frac{\chi_p}{\chi_k}\ep\cd k.
\eeq
Therefore, there is a one-to-one mapping between the electron's stationary phase (representing its classical trajectory) and the value of $\ep\cd k$ at which an ALP is emitted. This means the divergent integral in $\ep\cd k$ can be reinterpreted as an integration over the electron's phase $\varphi_{\ast}$ as it propagates through the background.
%
%
Writing the probabilities for the scalar and pseudo-scalar emissions as a probability per unit phase using $d(\ep\cd k)=\tfrac{\chi_k}{\chi_p}(m_e\xi)d\varphi_*$ we have
\begin{align} \label{lcfaP}
\frac{dP_+}{d\varphi_*}=&-\frac{g_{\phi e}^2 \xi}{8\pi}\int_{0}^{\infty}d\chi_k~\theta(\chi_p\!-\!\chi_k)\frac{1}{\chi_p^2}\Bigg[ \chi_k \left(\frac{\chi_k}{\chi_p(\chi_p\!-\!\chi_k)}\right)^{1/3} \text{Ai}^{\prime}(\Omega)-\left(2-\frac{1}{2}\delta^2\right)\text{Ai}_1(\Omega)\Bigg]	\non\\
\frac{dP_-}{d\varphi_*}=&-\frac{g_{\phi e}^2 \xi}{8\pi}\int_{0}^{\infty}d\chi_k~\theta(\chi_p\!-\!\chi_k)\frac{1}{\chi_p^2}\Bigg[ \chi_k \left(\frac{\chi_k}{\chi_p(\chi_p\!-\!\chi_k)}\right)^{1/3} \text{Ai}^{\prime}(\Omega)+\frac{1}{2}\delta^2\text{Ai}_1(\Omega)\Bigg]	
\end{align}
where the difference in the two lies in the pre-factor of the $\text{Ai}_1$ term.
To calculate the probability of emission in a non-trivial external field we then use the Locally Constant Field Approximation (LCFA) and make the replacement
\beq
\xi=\xi_0 g^\prime(\varphi_*)~~~\text{and}~~~\chi_q(\varphi_*)=\chi_{q,0}g^\prime(\varphi_*)
\eeq
where $g^\prime(\varphi_*)$ is the profile of the electric field.

In Appendix \ref{angular} we discuss the transformation from $d(\ep\cd k)dk^-$ to polar coordinates and the consequences for light-front invariants.
The same transformation can also be used here to obtain information on the angular distribution of the emitted ALPs in a non-trivial external field background.
Using Eq.~\ref{psistar} we can write
\beq\label{angP}
m_e\frac{dP_\pm}{d(\ep\cd k)d\chi_k}=-g^\prime(\ep\cd k)\frac{g_{\phi e}^2}{8\pi}\frac{\chi_p}{\chi_k}\left[\theta(k^--p^-)\ldots\right]
\eeq
where $g^\prime(\ep\cd k)$ is simply the external electric field profile written in terms of the ALP momenta and the ellipsis corresponds to the integrands in Eq. \ref{lcfaP} written with the replacement $\chi_{p,k}\rightarrow\chi_{p,k}(\varphi_*)$.
Note that the Eq.~\ref{angP} does not explicitly depend on the non-linearity parameter $\xi$ as this only enters through the $\chi$-parameters, also the ratios $\tfrac{\chi_k}{\chi_p}=\tfrac{k^-}{p^-}$ remain independent of both $\xi$ and $g(\varphi_*)$.

\subsection{Yield distributions in a constant field background}  \label{yieldCF}

\begin{figure}[H]
\centering
  \includegraphics[width=120mm]{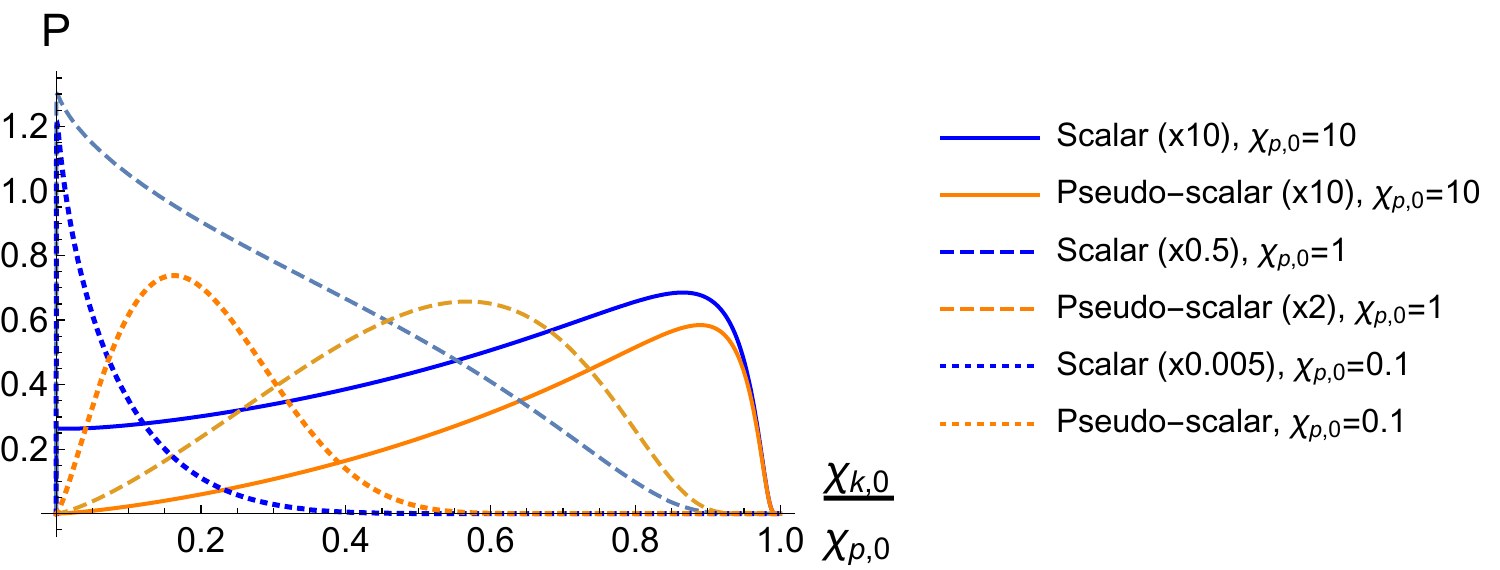}
\caption{The relationship between the differential yield and the $\chi_k$ parameter is plotted for different seed electron energies with $\xi_0=100$ and $g_{\phi e}=1$\protect\footnotemark.\label{cf1}}
\end{figure}

\noindent We can study properties of the yield distribution for scalars and pseudo-scalars in a constant field by taking the external field profile to be constant over some finite distance, i.e.
\beq
g(\varphi_*)=\Theta(L+\varphi_*)\Theta(L-\varphi_*)
\eeq 
with $\Theta(x)$ being the Heaviside step function and $L$ being some finite phase.
The yield distribution in $\chi_k$ remains constant with $\varphi_*$ thus we can sample the distribution at one point to examine its behaviour, this is shown in Figure \ref{cf1} where we have set $\gp=1$.
In this figure we display the yield distributions in $\chi_{k}$ for various seed electron energies, where the probabilities have been re-scaled for purposes of comparison.
There is a clear difference between the scalar and pseudo-scalar $\chi_k$ distributions.
The distribution for pseudo-scalar production is peaked away from zero for all values of $\chi_{p,0}$, with the maximum of the distribution moving closer to $\chi_{k,0}=\chi_{p,0}$ for larger $\chi_{p,0}$.
For $\chi_{p,0}\lesssim 1$ the distribution in $\chi_k$ for scalar production is peaked at zero, whereas for larger values of $\chi_{p,0}$ the distribution becomes peaked away from zero and begins to resemble the distribution for pseudo-scalar production.
\footnotetext{For the plots in this section we take $\xi_0=100$ and $g_{\phi e}=1$ unless otherwise stated.}

\subsection{Yield distributions in a Gaussian background}

\begin{figure}[H]
\centering
\hspace{-7mm}
\subfloat[$\chi_{p,0}=0.1$]{
\raisebox{-0.6ex}{
  \includegraphics[scale=0.45]{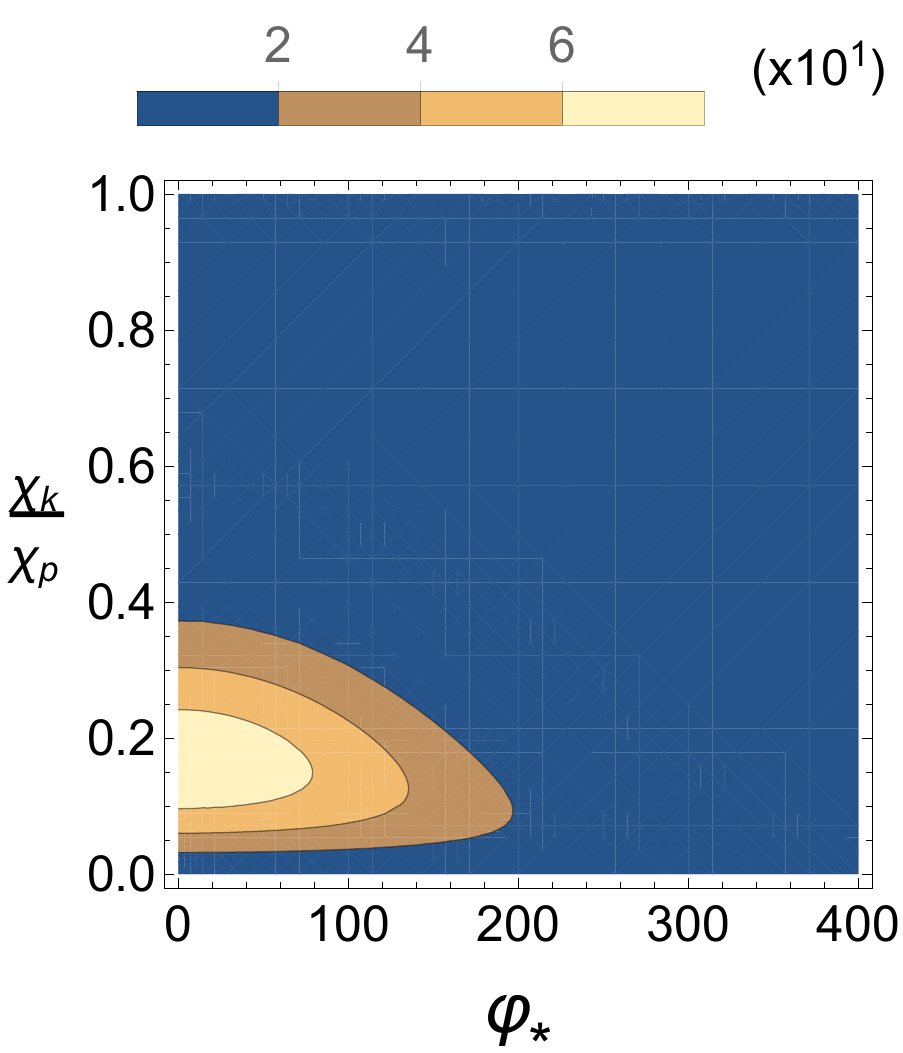}}
}
\subfloat[$\chi_{p,0}=0.4$]{
 \raisebox{0.3ex}{
  \includegraphics[scale=0.375]{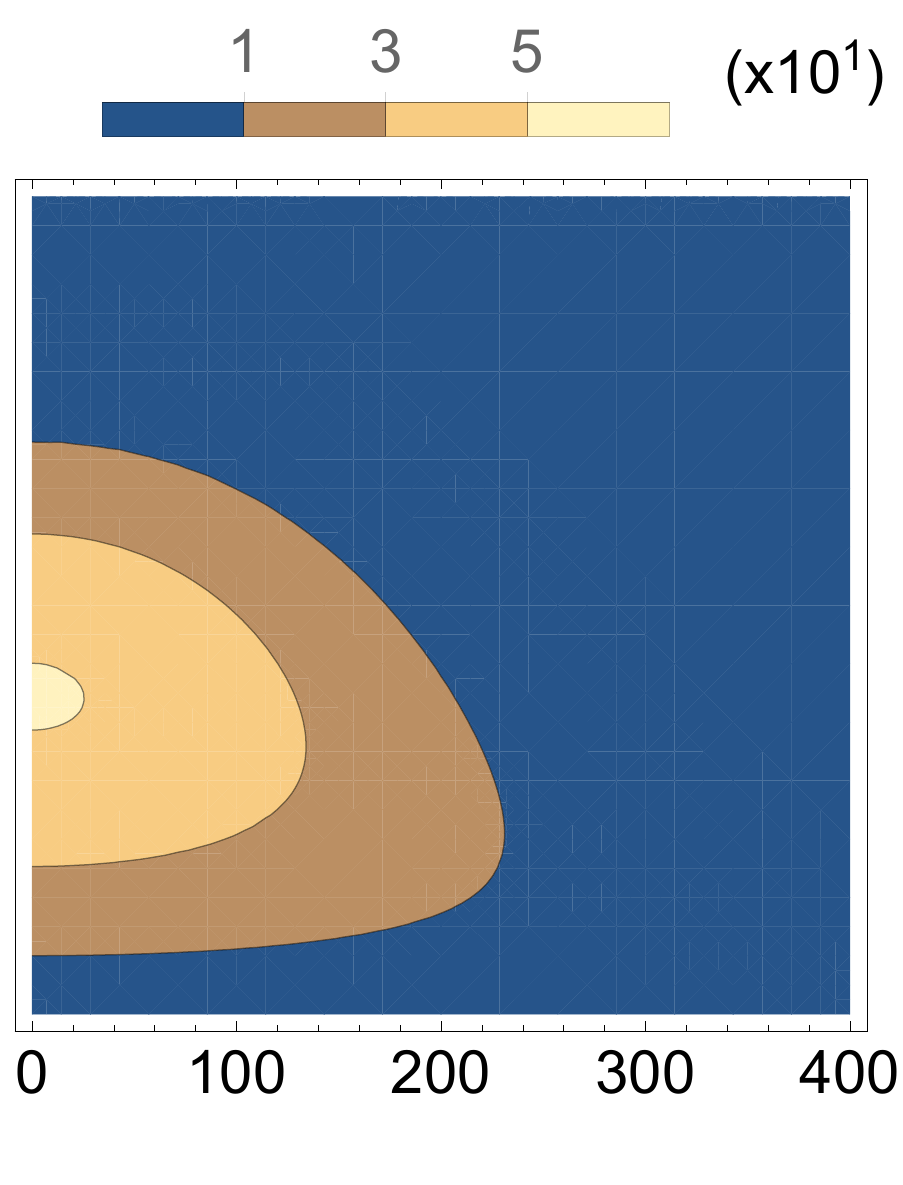}}
}
\subfloat[$\chi_{p,0}=1$]{
\raisebox{0.3ex}{
 \includegraphics[scale=0.375]{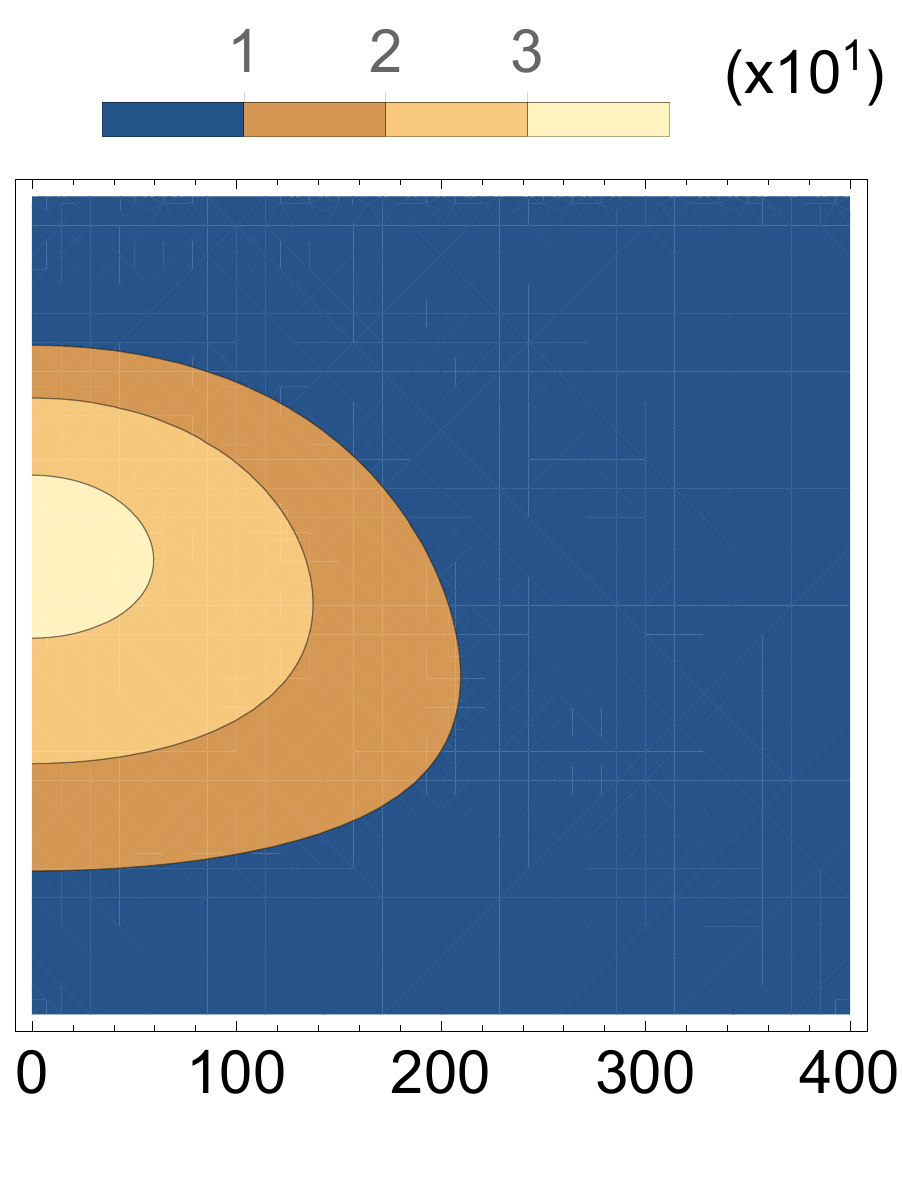}}
}
\subfloat[$\chi_{p,0}=10$]{
 \raisebox{0.3ex}{
  \includegraphics[scale=0.375]{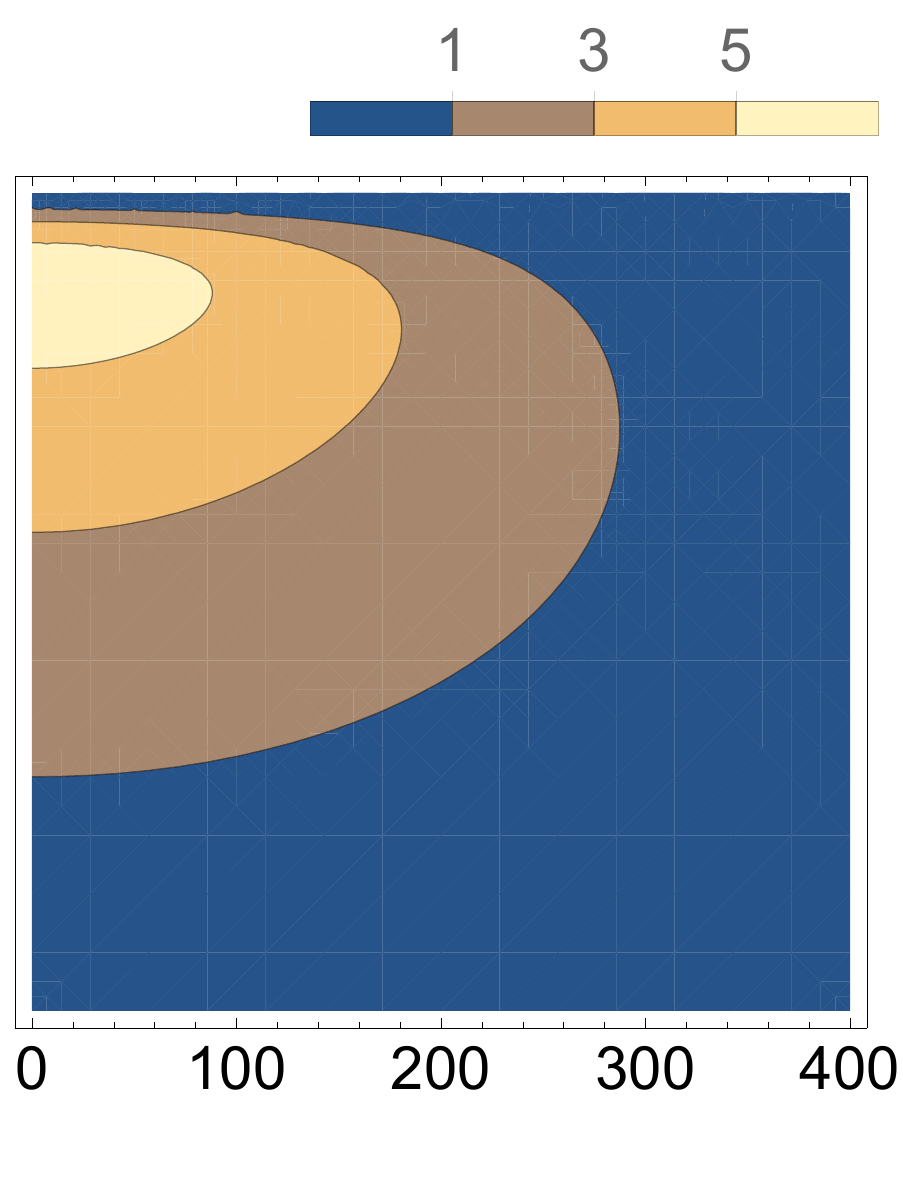}}
}
\caption{The pseudo-scalar yield, $P_-$, is plotted as a function of $\chi_k/\chi_p$ and the phase $\varphi_*$ in a Gaussian background with a pulse duration of $100$ fs and $m_{\phi}=0$. \label{Gpseudoscalar1}}
\end{figure}

\begin{figure}[H]
\centering
\hspace{-7mm}
\subfloat[$\chi_{p,0}=0.1$]{
\raisebox{-0.6ex}{
  \includegraphics[scale=0.45]{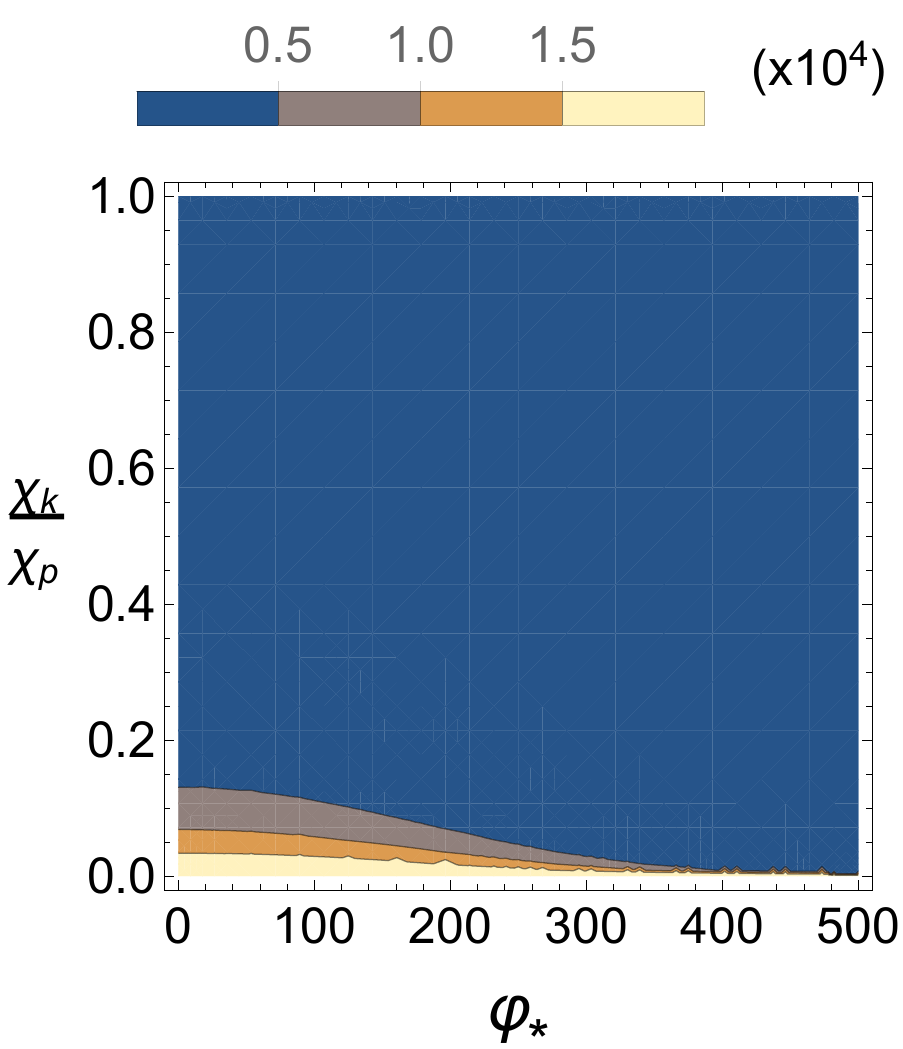}}
}
\subfloat[$\chi_{p,0}=0.4$]{
 \raisebox{0.3ex}{
  \includegraphics[scale=0.375]{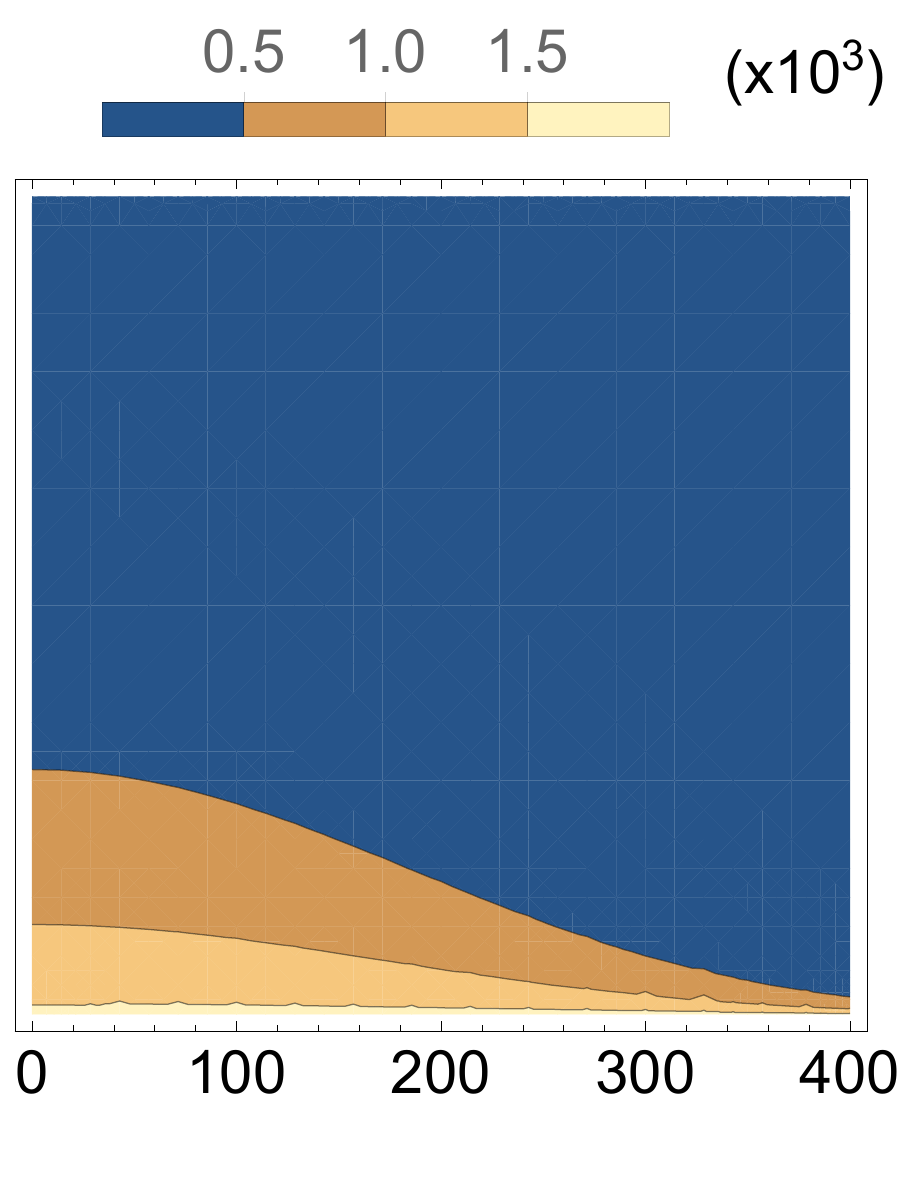}}
}
\subfloat[$\chi_{p,0}=1$]{
\raisebox{0.3ex}{
 \includegraphics[scale=0.375]{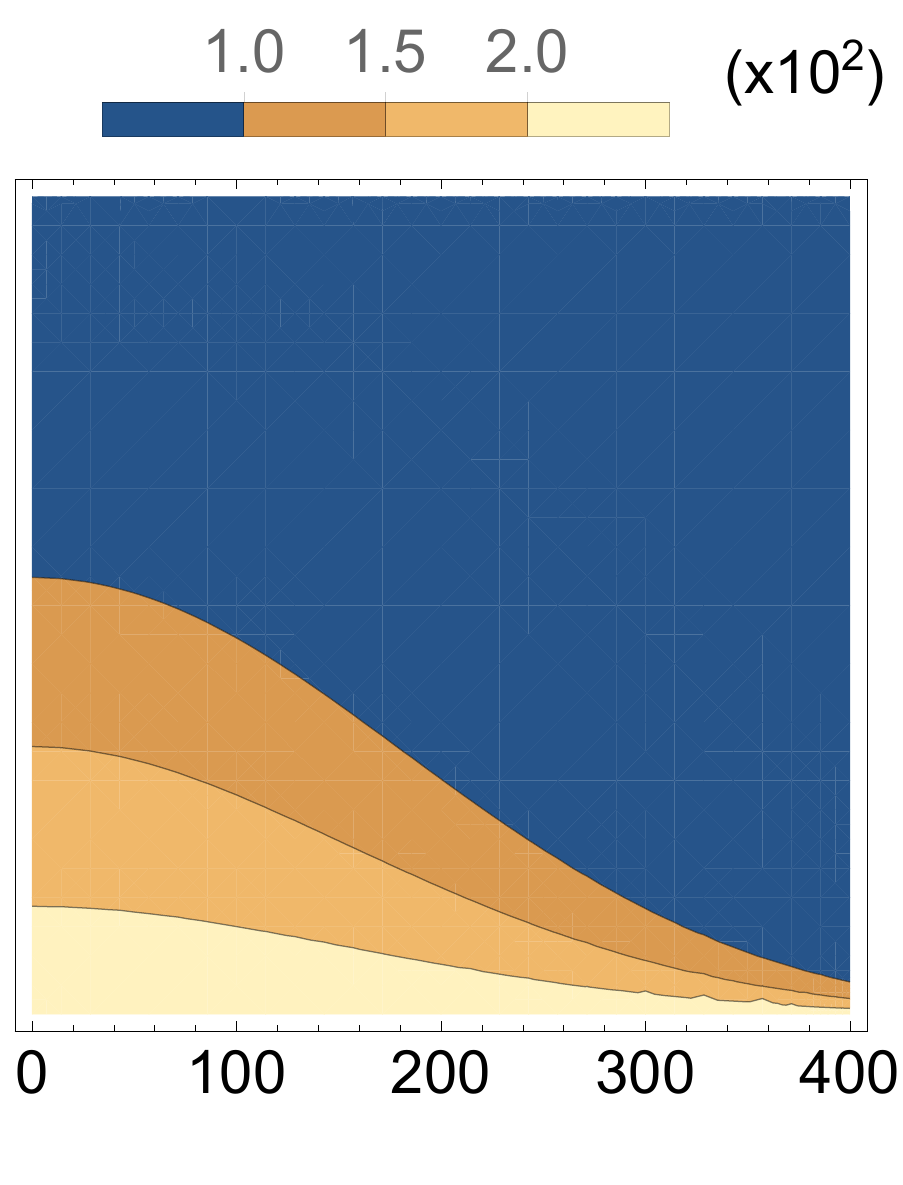}}
}
\subfloat[$\chi_{p,0}=10$]{
 \raisebox{0.3ex}{
  \includegraphics[scale=0.375]{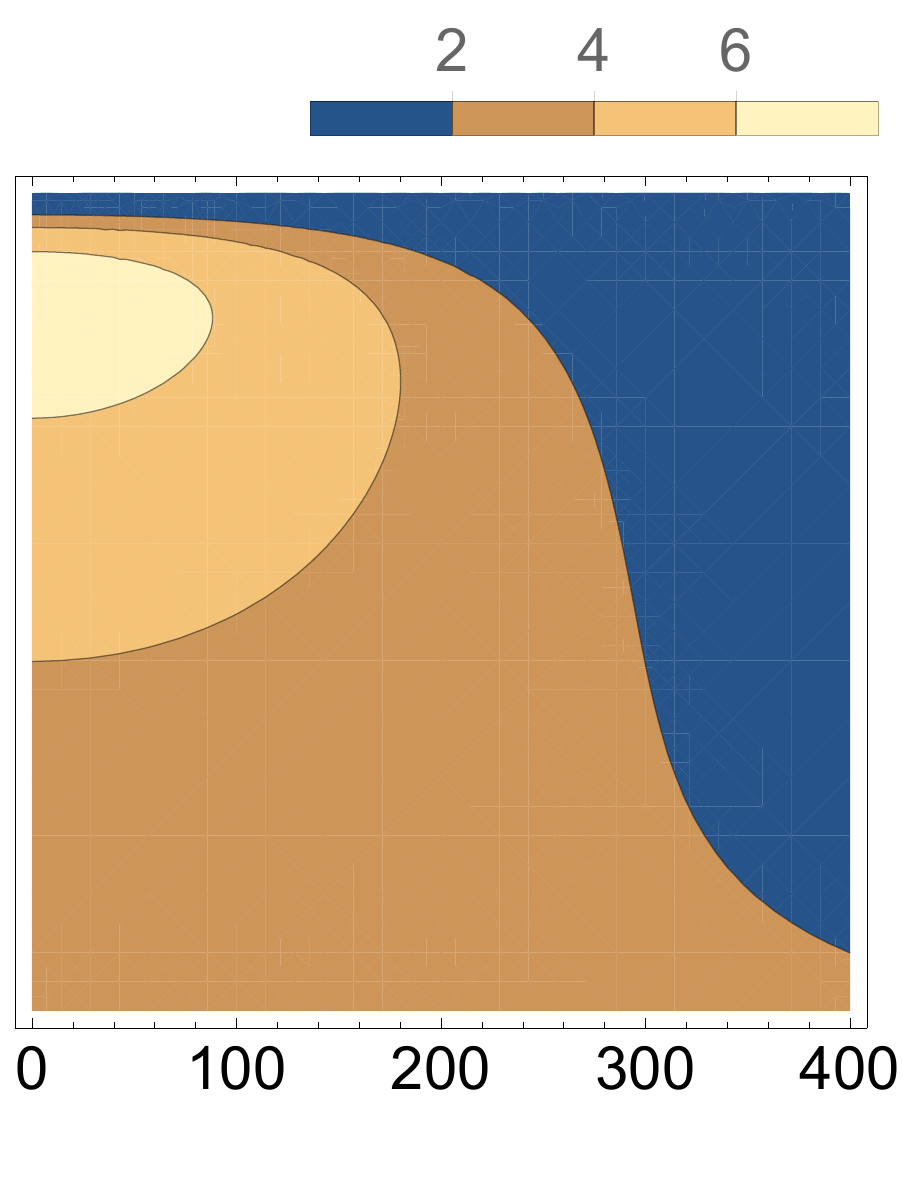}}
}
\caption{The scalar yield, $P_+$, is plotted as a function of $\chi_k/\chi_p$ and the phase $\varphi_*$ in a Gaussian background with a pulse duration of $100$ fs and $m_{\phi}=0$. \label{Gscalar1}}
\end{figure}

\noindent We now study the emission of pseudo-scalars and scalars from an electron in an external field with a Gaussian profile described by
\beq
g(\varphi_*)=e^{-\left(\frac{\varphi_*}{\Phi}\right)^2}
\eeq 
with $\Phi$ being the pulse duration in units of inverse $\kappa_0$.
We choose the duration for the high-intensity laser pulse to be $100$~fs throughout this chapter (corresponding to $\Phi \approx 300$).
The distribution in $\chi_{k,0}$ now has a non-trivial dependence on the phase $\varphi_*$.
To show this we have plotted the pseudo-scalar and scalar yields for various seed electron energies in Figures \ref{Gpseudoscalar1} and \ref{Gscalar1}, respectively, where we have set $\gp=1$.
In the pseudo-scalar case we see that the distribution is localised at a point which for low seed electron energies is at $\chi_{k,0}\ll\chi_{p,0}$, but for larger seed electron energies moves towards $\chi_{k,0}\simeq\chi_{p,0}$.
This is in contrast to the scalar case in which the distribution is always localised around $\chi_{k,0}\simeq0$, where larger seed electron energies increase the spread of the distribution towards $\chi_{k,0}\simeq\chi_{p,0}$.
Similar behaviour can also be seen in the constant field case, Figure \ref{cf1}.
Note that these distributions are symmetric around $\varphi_*=0$, which corresponds to the central peak of the Gaussian profile in $g^\prime(\varphi_*)$.

\begin{figure}[H]
\centering
\hspace*{-7mm}
\subfloat[$\chi_{p,0}=0.1$]{
\raisebox{-0.6ex}{
  \includegraphics[scale=0.53]{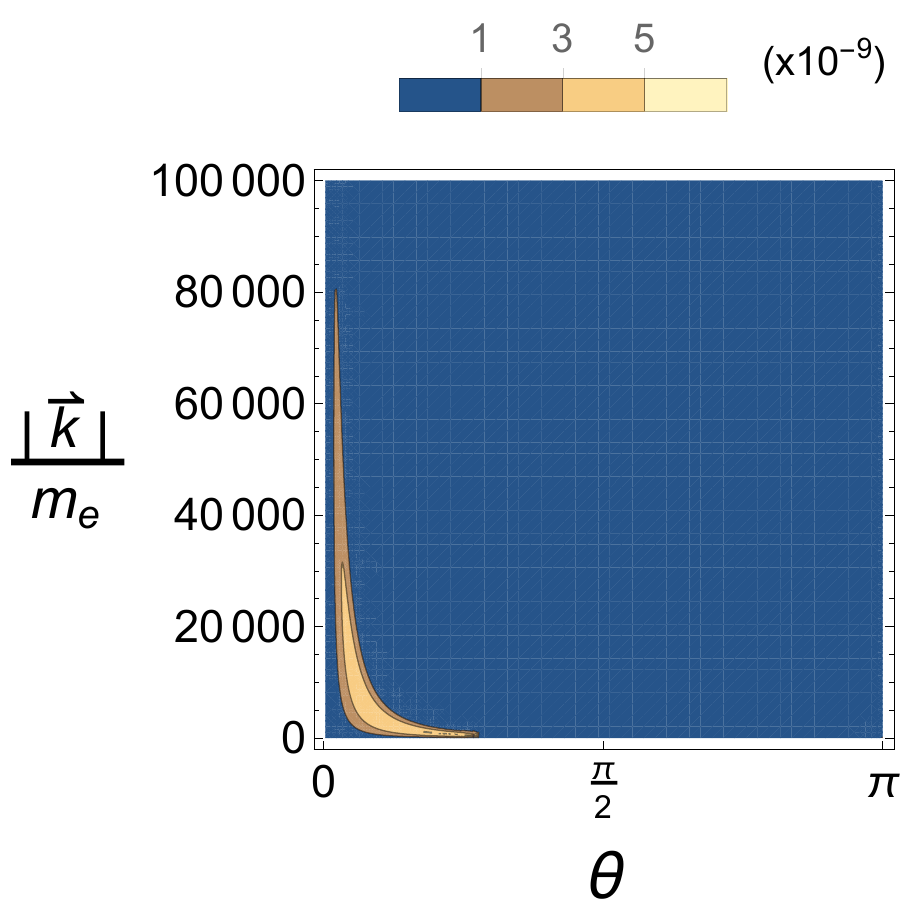}}
}
\subfloat[$\chi_{p,0}=1$]{
 \raisebox{0.8ex}{
  \includegraphics[scale=0.345]{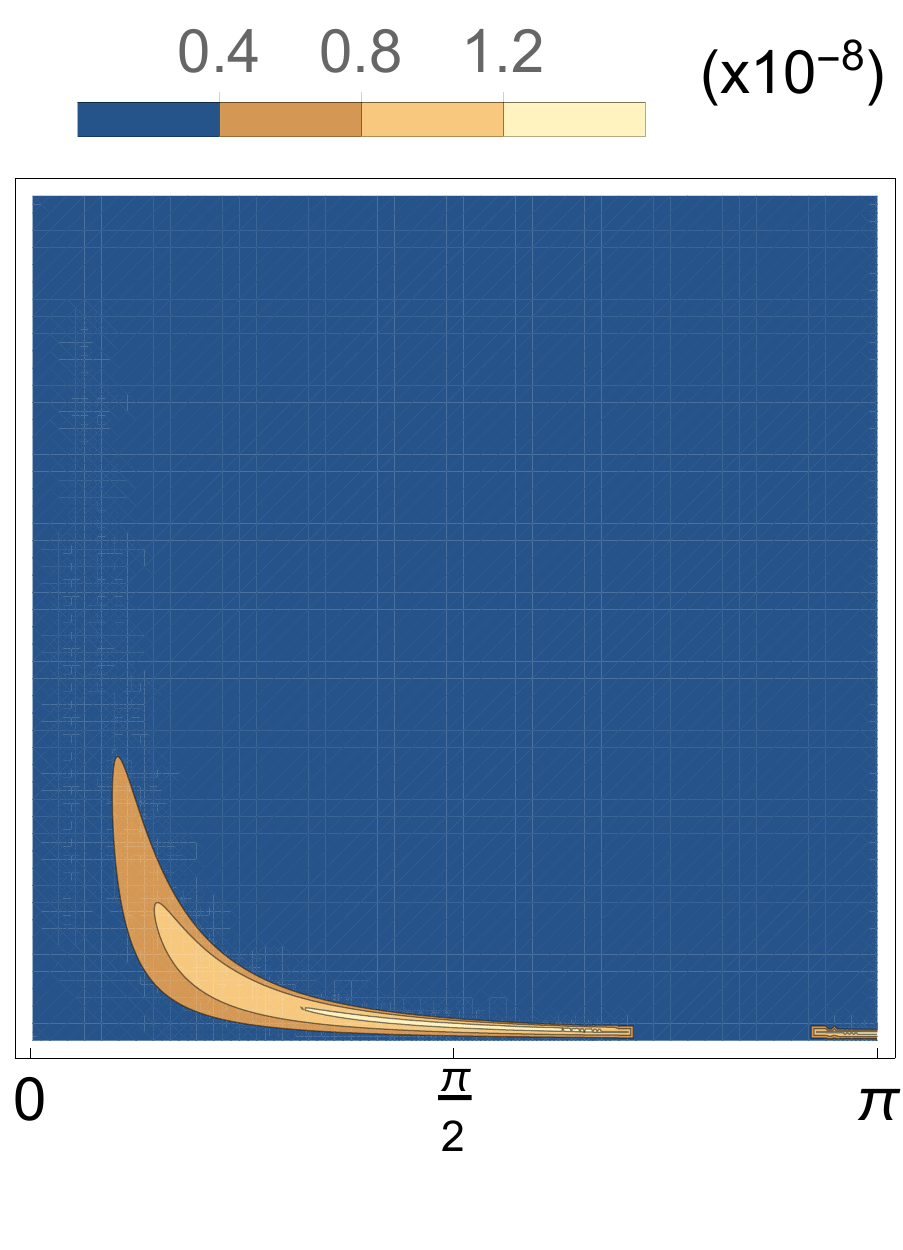}}
}
\subfloat[$\chi_{p,0}=10$]{
\raisebox{0.8ex}{
 \includegraphics[scale=0.345]{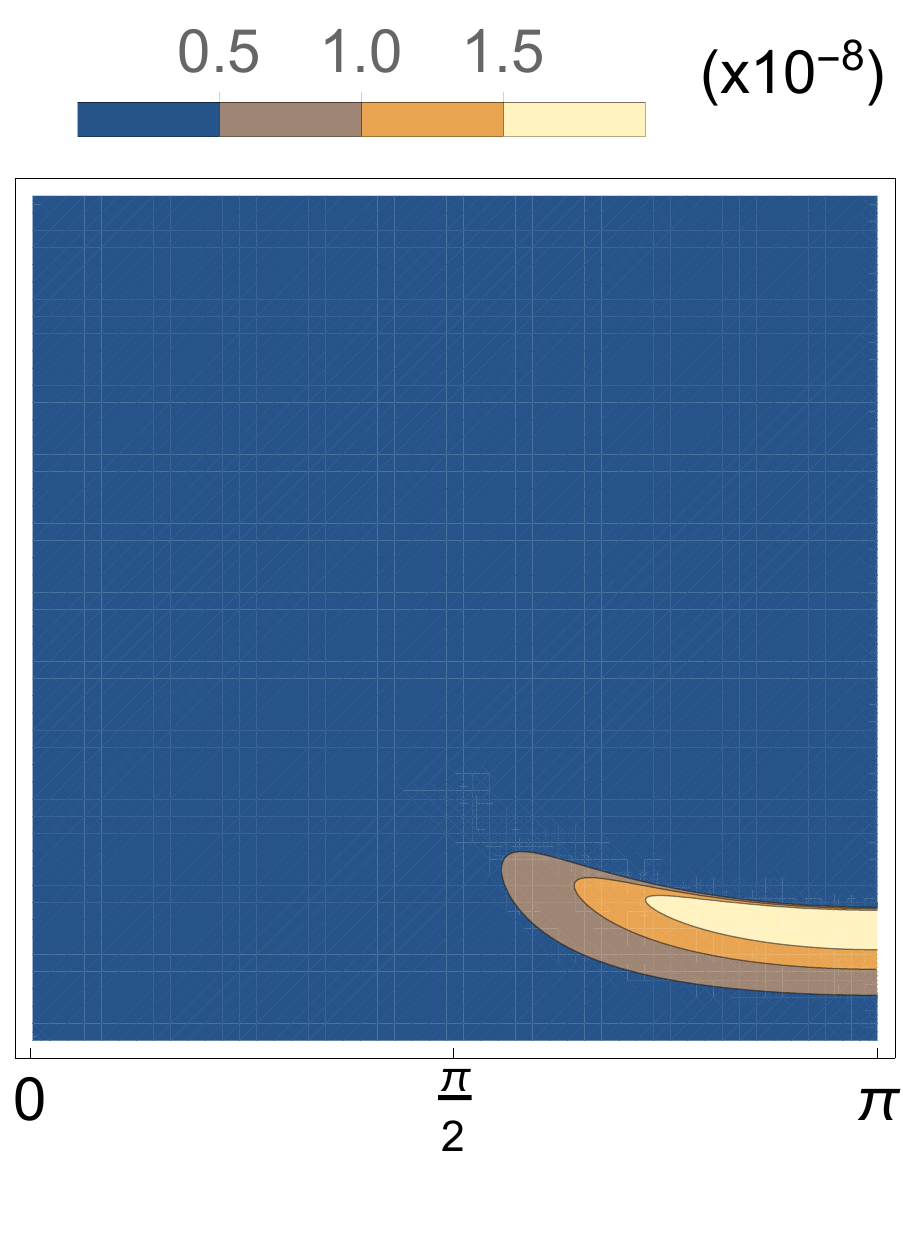}}
}
\subfloat[$\chi_{p,0}=40$]{
 \raisebox{0.8ex}{
  \includegraphics[scale=0.345]{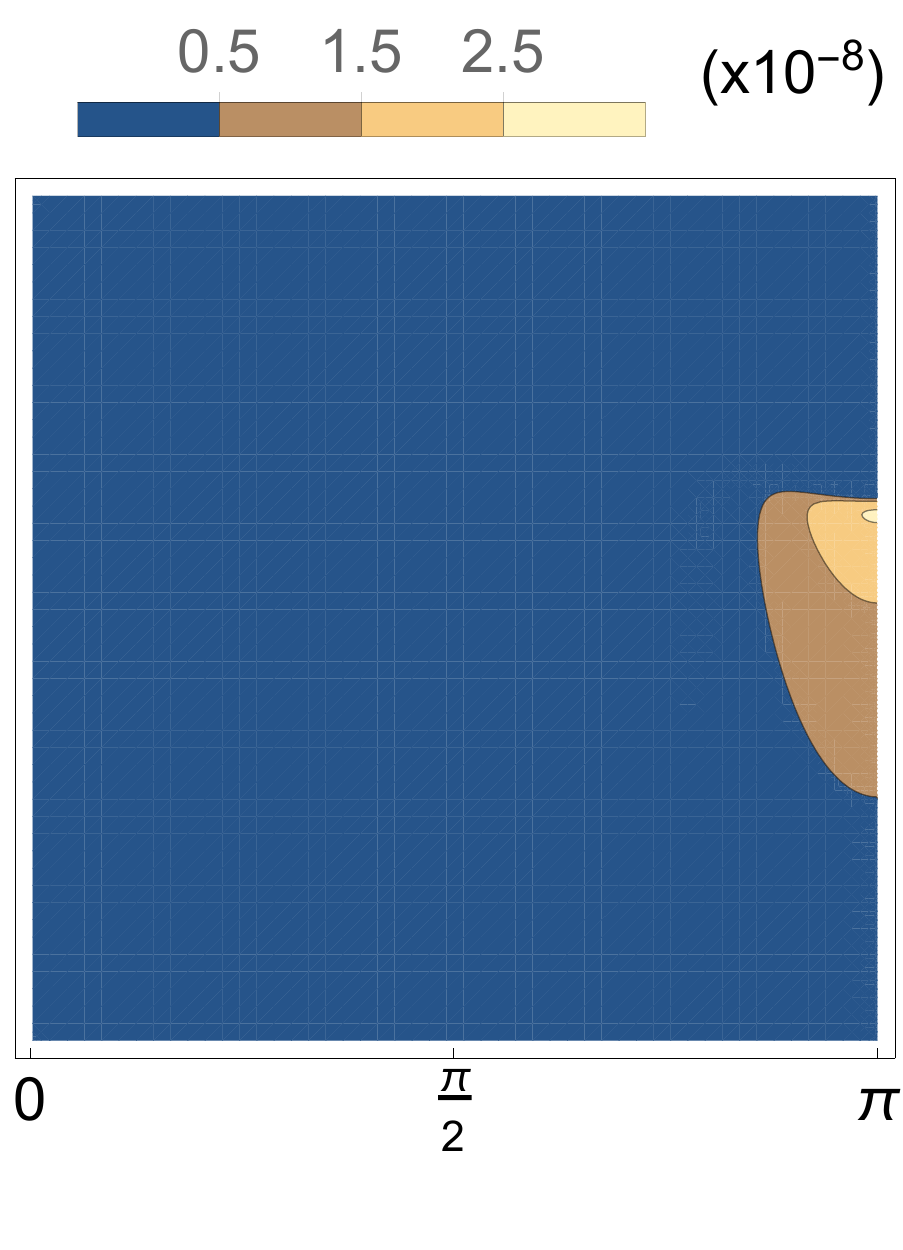}}
}
\caption{The pseudo-scalar yield, $P_-$, is plotted as a function of the angle and momenta of the emitted ALP in a Gaussian background with $m_{\phi}=0$.\label{Gpseudoscalar2}}
\end{figure}

\begin{figure}[H]
\centering
\hspace*{-7mm}
\subfloat[$\chi_{p,0}=0.1$]{
\raisebox{-0.6ex}{
  \includegraphics[scale=0.53]{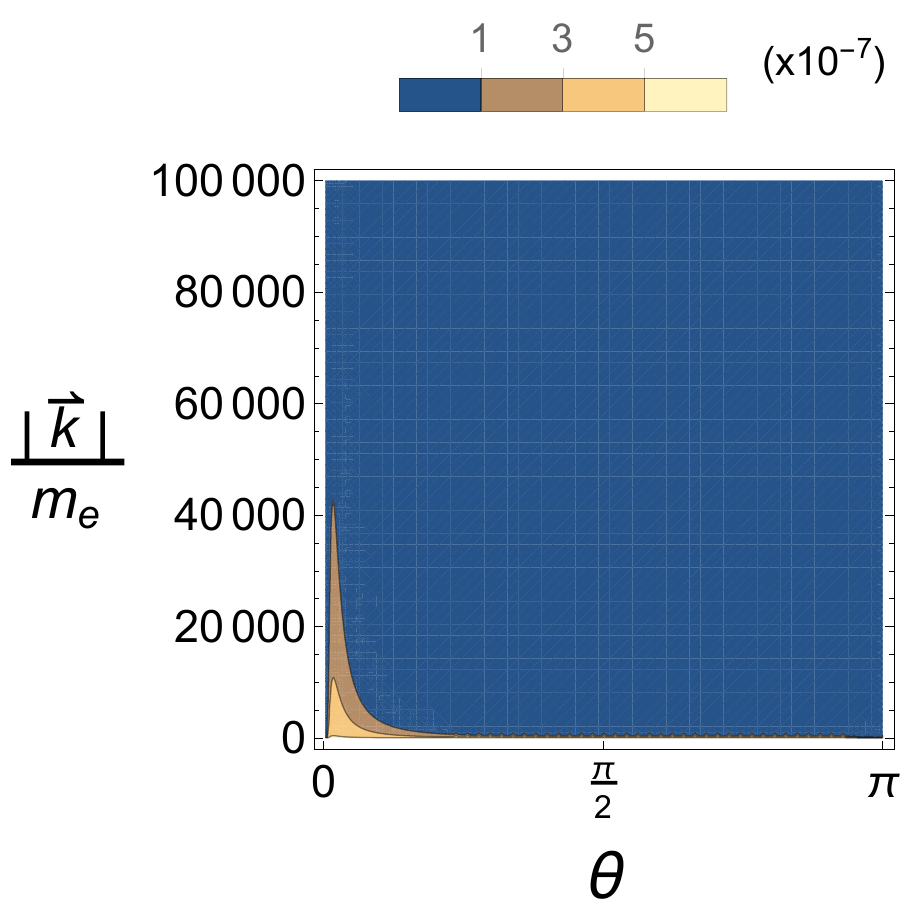}}
}
\subfloat[$\chi_{p,0}=1$]{
 \raisebox{0.8ex}{
  \includegraphics[scale=0.345]{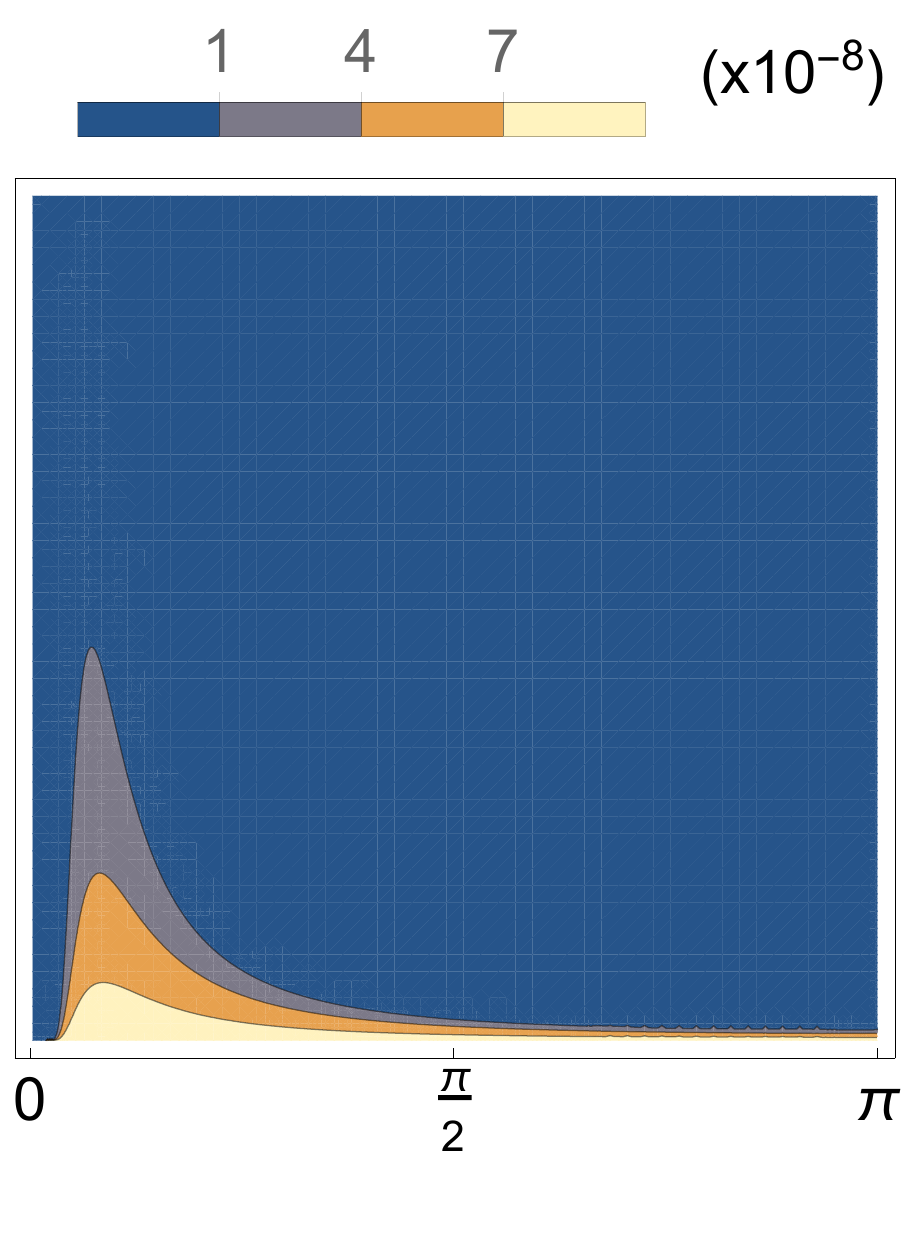}}
}
\subfloat[$\chi_{p,0}=10$]{
\raisebox{0.8ex}{
 \includegraphics[scale=0.345]{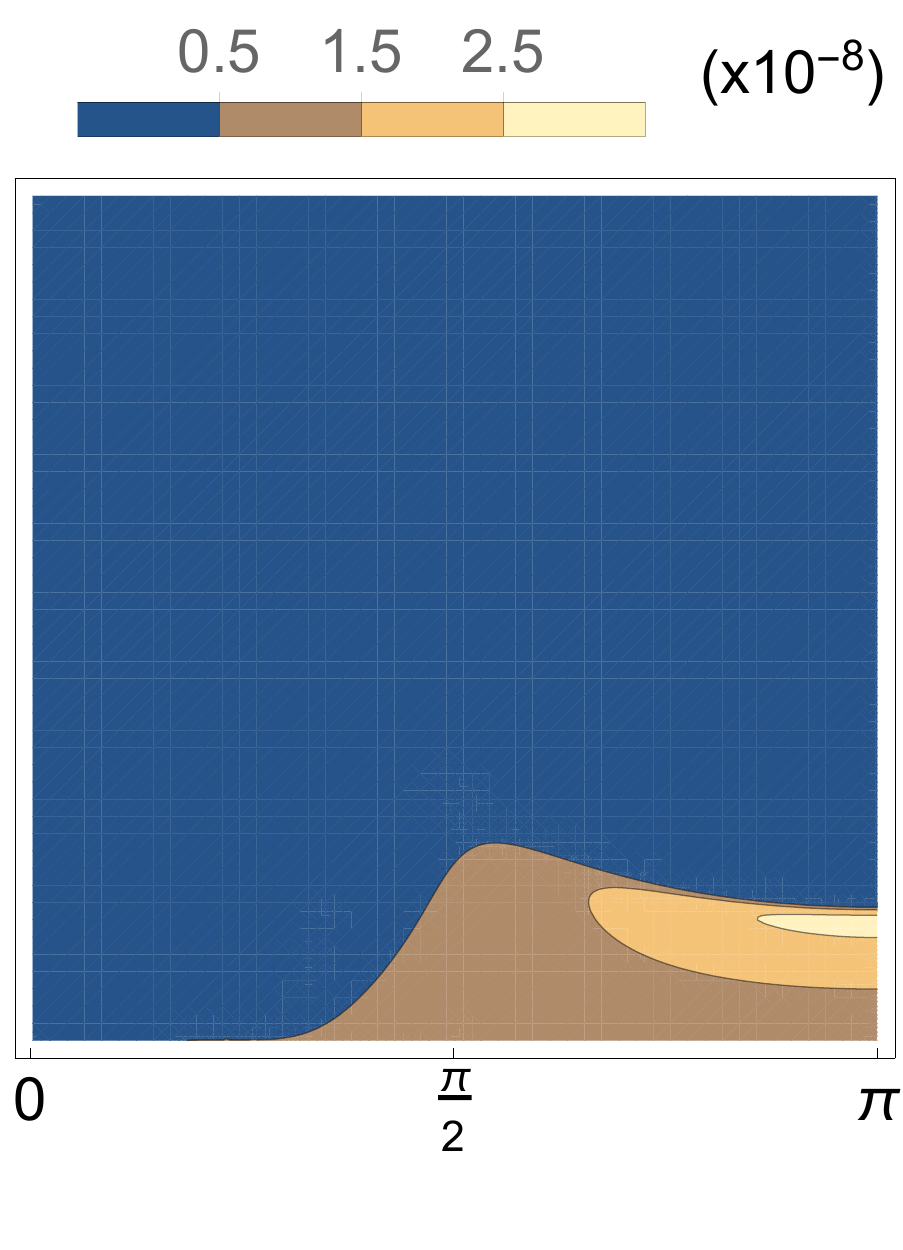}}
}
\subfloat[$\chi_{p,0}=40$]{
 \raisebox{0.8ex}{
  \includegraphics[scale=0.345]{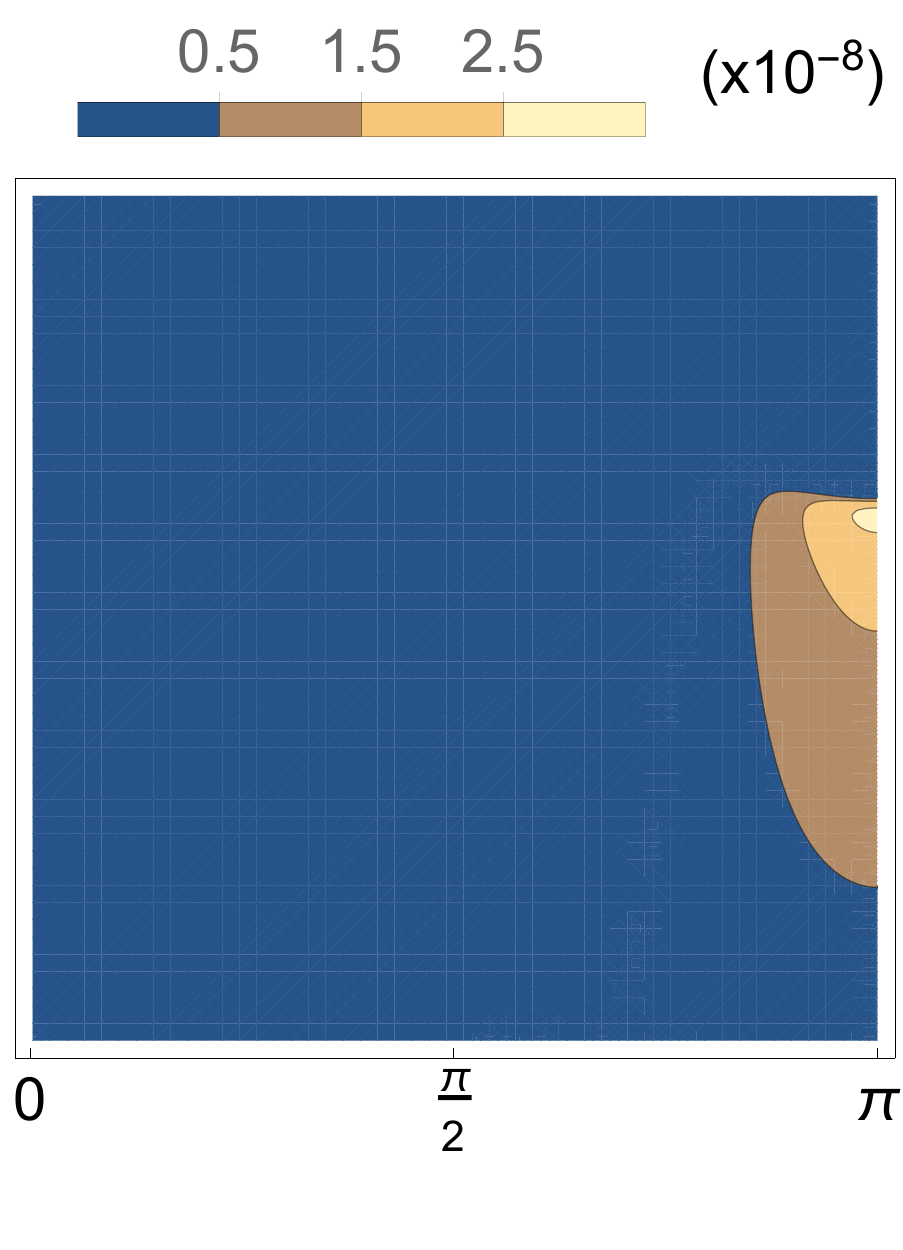}}
}
\caption{The scalar yield, $P_+$, is plotted as a function of the angle and momenta of the emitted ALP in a Gaussian background with $m_{\phi}=0$.\label{Gscalar2}}
\end{figure}

In Figures \ref{Gpseudoscalar2} and \ref{Gscalar2} we show, using Eq. \ref{polar}, how this translates to distributions in the polar angle and energy of the emitted ALP in lab frame coordinates.
For both the pseudo-scalar and scalar yields the higher energy particles are emitted at smaller polar angles for small seed electron energies.
Larger seed electron energies results in the ALPs being emitted at larger angles, becoming parallel with the direction of propagation of incoming electrons for very large seed electron energies.
These distributions are similar to the constant-crossed field case in that the scalar yield is peaked at $|\vec{k}|\simeq0$ while the pseudo-scalar yield is peaked at a non-zero $|\vec{k}|$ determined by the seed electron energy.

\subsection{Total yields and ALP mass dependence}

\noindent Thus far in this section we have taken the ALP to be massless, i.e. $\delta=0$.
For the mass effects to become significant in the high-intensity regime we require $\delta\gtrsim 0.1$.
This contrasts with the results we found in the low intensity external field in Section \ref{perturbative}, where the range of masses probed by the interaction was limited by the energy of the photons in the background field.
In this section we will study what happens to the differential yield and the total yield when we allow for sizeable ALP masses.

\begin{figure}[H]
\centering
\includegraphics[width=160mm]{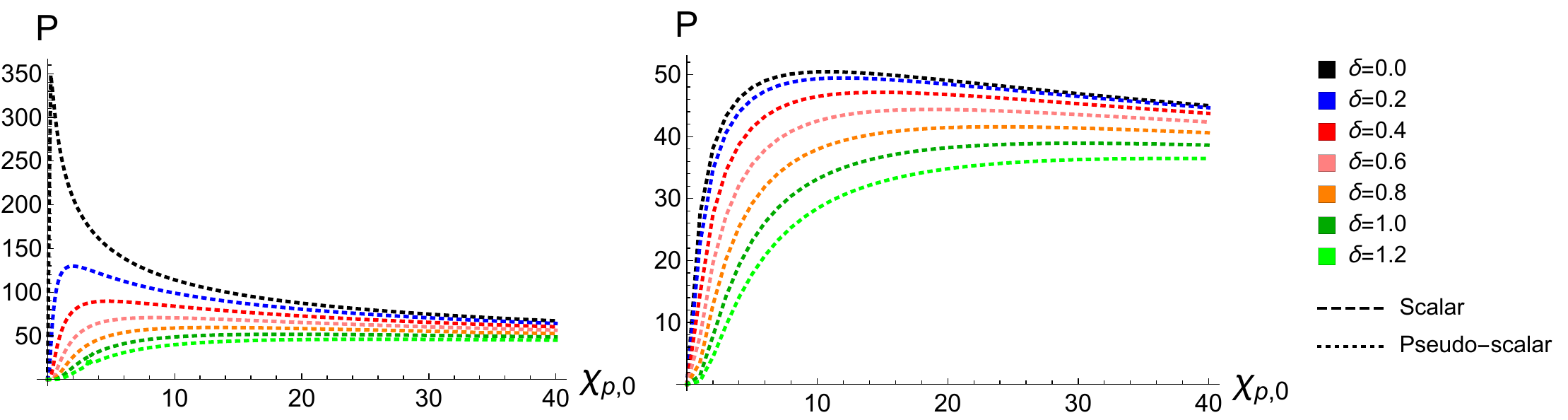}
\caption{The pseudo-scalar and scalar yields, $P_\pm$, are plotted as a function of $\chi_{p,0}$ for various ALP masses, with $m_{\phi}=\delta m_e$, in a Gaussian background.\label{TYmassD}}
\end{figure}

We start by plotting the total integrated yield for a Gaussian background field against the seed electron $\chi_{p,0}$ for various values of $\delta$ in Figure \ref{TYmassD}.
The analogous plot for a constant field background has the same features but at a different magnitude, scaled by how long the electron is taken to interact with the constant field (which is of course, formally infinite).
In Figure \ref{TYmassD} we have allowed for a much larger range in $\chi_{p,0}$ than in previous plots, this is only done because the full picture of the effects related to a non-zero ALP mass on the total yield only become apparent at these larger values of $\chi_{p,0}$.

\begin{figure}[H]
\centering\hspace{-7mm}
\subfloat[$P_-,~\delta=0$]{
\raisebox{-0.6ex}{
  \includegraphics[scale=0.45]{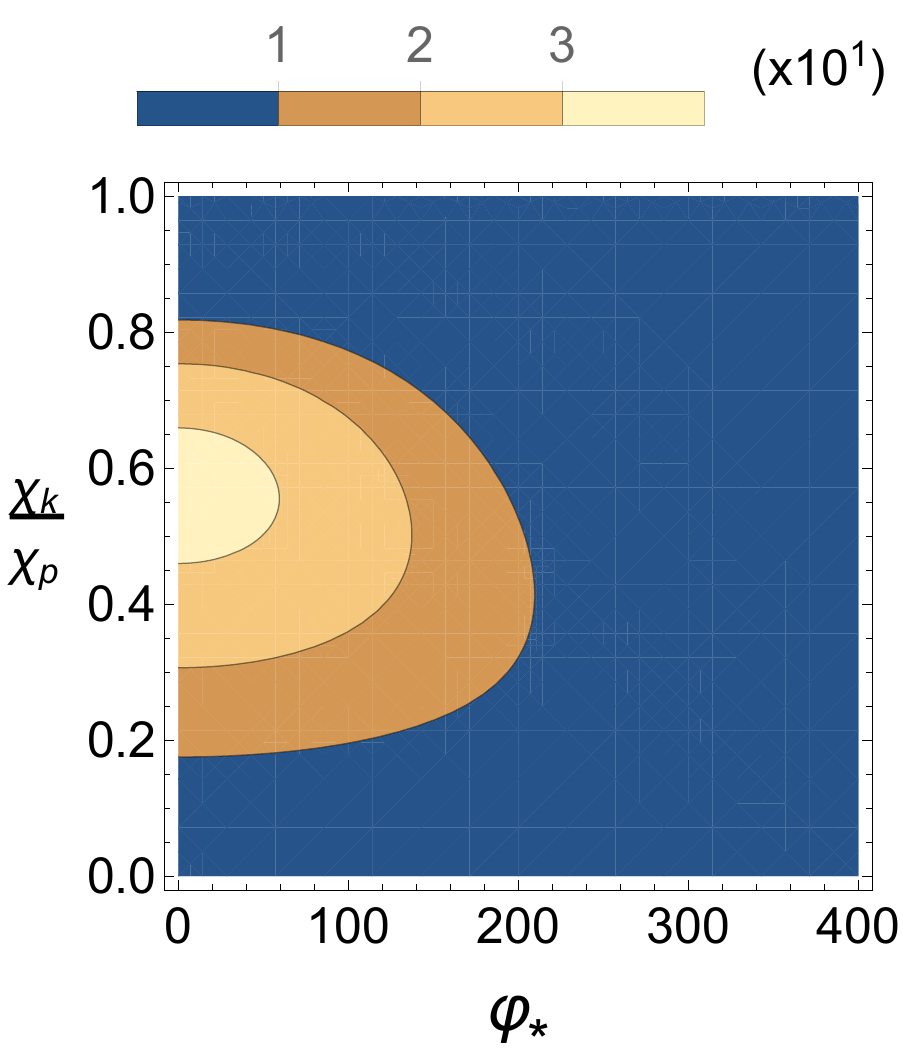}}
}
\subfloat[$P_-,~\delta=0.6$]{
 \raisebox{0.1ex}{
  \includegraphics[scale=0.375]{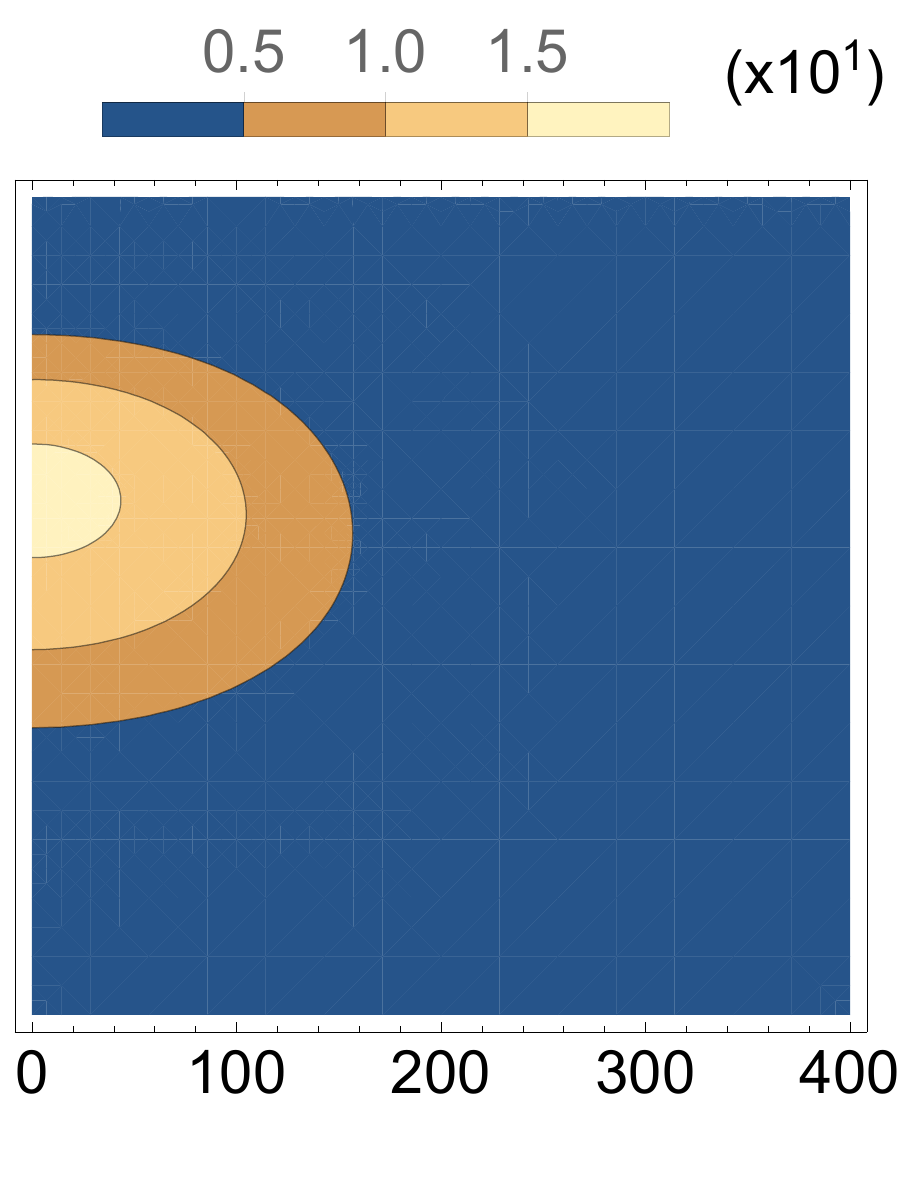}}
}
\subfloat[$P_+,~\delta=0$]{
\raisebox{0.1ex}{
 \includegraphics[scale=0.375]{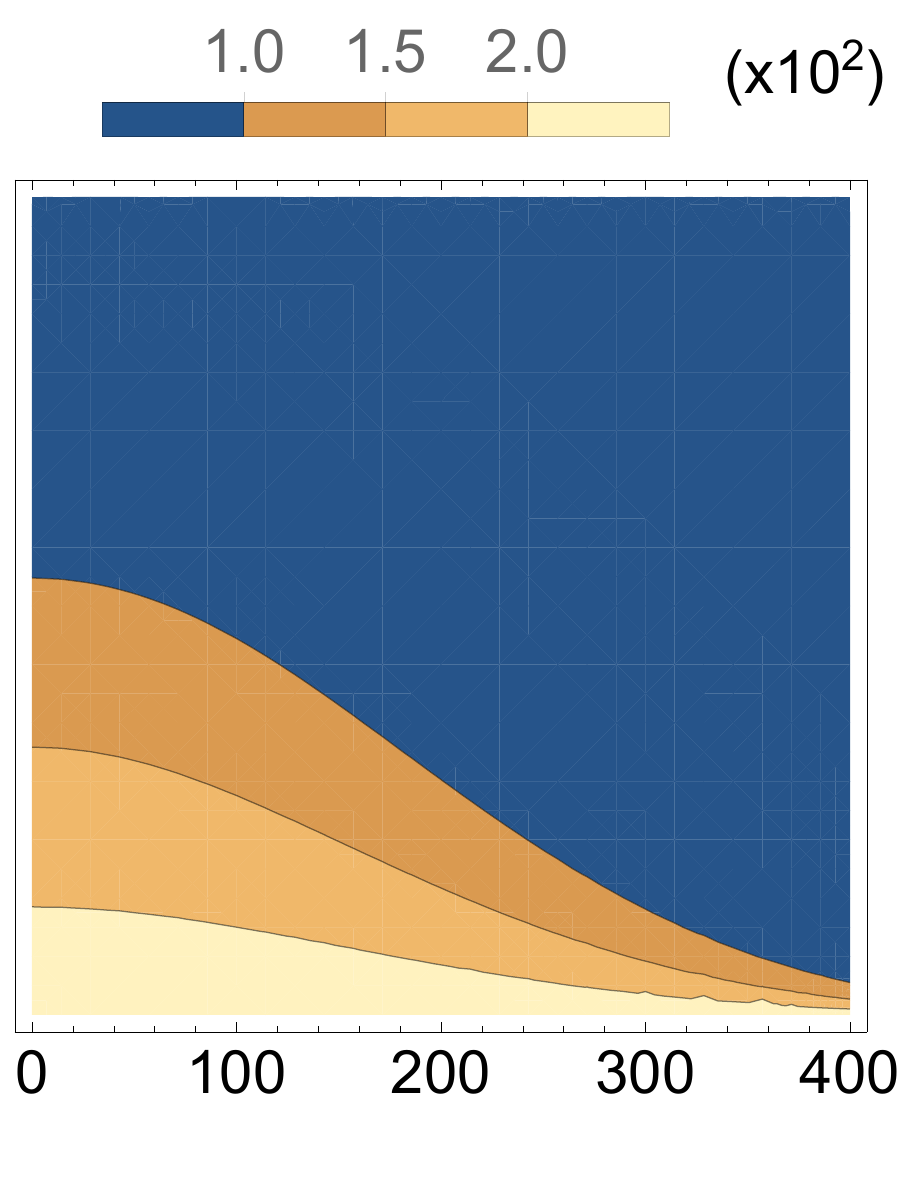}}
}
\subfloat[$P_+,~\delta=0.6$]{
 \raisebox{0.1ex}{
  \includegraphics[scale=0.375]{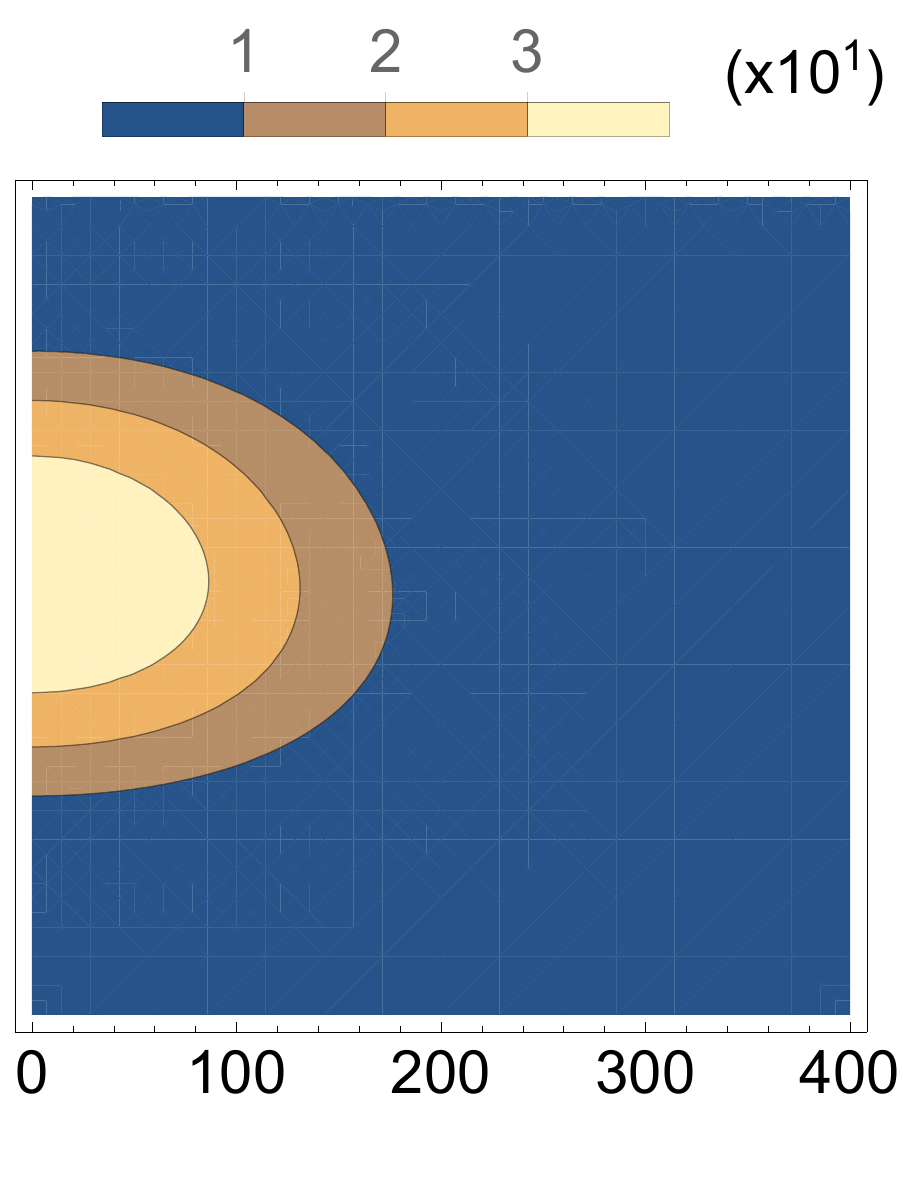}}
}
\caption{The pseudo-scalar (left) and scalar (right) yields, $P_\pm$, are plotted as a function of the $\chi_k/\chi_p$ and the phase $\varphi_*$ in a Gaussian background for two different ALP masses.  Also, $\chi_{p,0}$ has been fixed to $1$.  \label{cpd}}
\end{figure}

The sharp peak for the massless scalar can be resolved in the logarithmic plot in Figure \ref{TYd0log} below.
We can see that there is a steep exponential increase in the production rate as $\chi_{p,0}$ is increased to $\chi_{p,0}\sim0.3$ followed by a gentler exponential decrease.
It is also interesting to see the effects of $\delta\neq0$ on the $\chi_k$ distribution of the emitted ALP in a constant field, we have plotted this in Figure \ref{cpd} where $\chi_{p,0}$ has been fixed to $1$.
\begin{figure}[H]
\centering
\includegraphics[width=60mm]{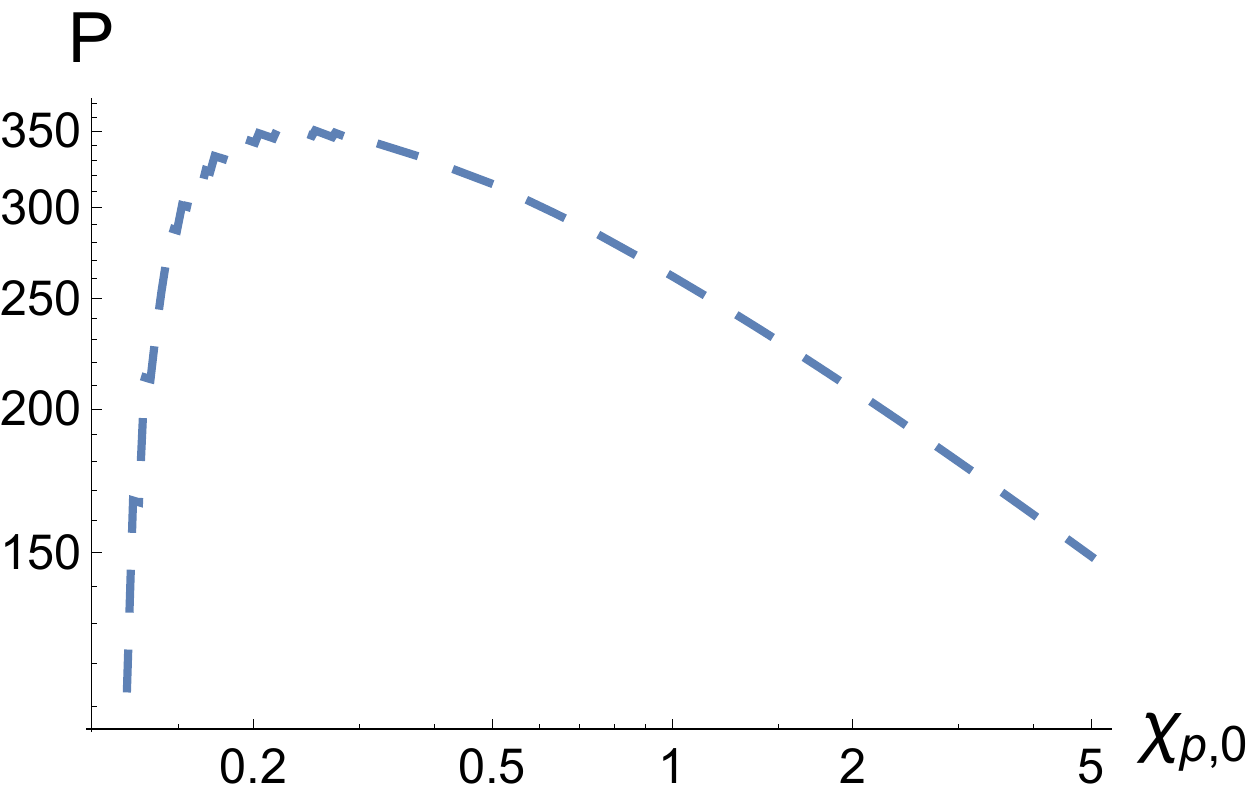}
\caption{The scalar yield, $P_+$ is plotted as a function of $\chi_{p,0}$ for $\delta=0$ in a Gaussian background.\label{TYd0log}}
\end{figure}

\begin{figure}[H]
\centering
\hspace*{-7mm}
\subfloat[$P_-,~\delta=0$]{
\raisebox{-0.6ex}{
  \includegraphics[scale=0.515]{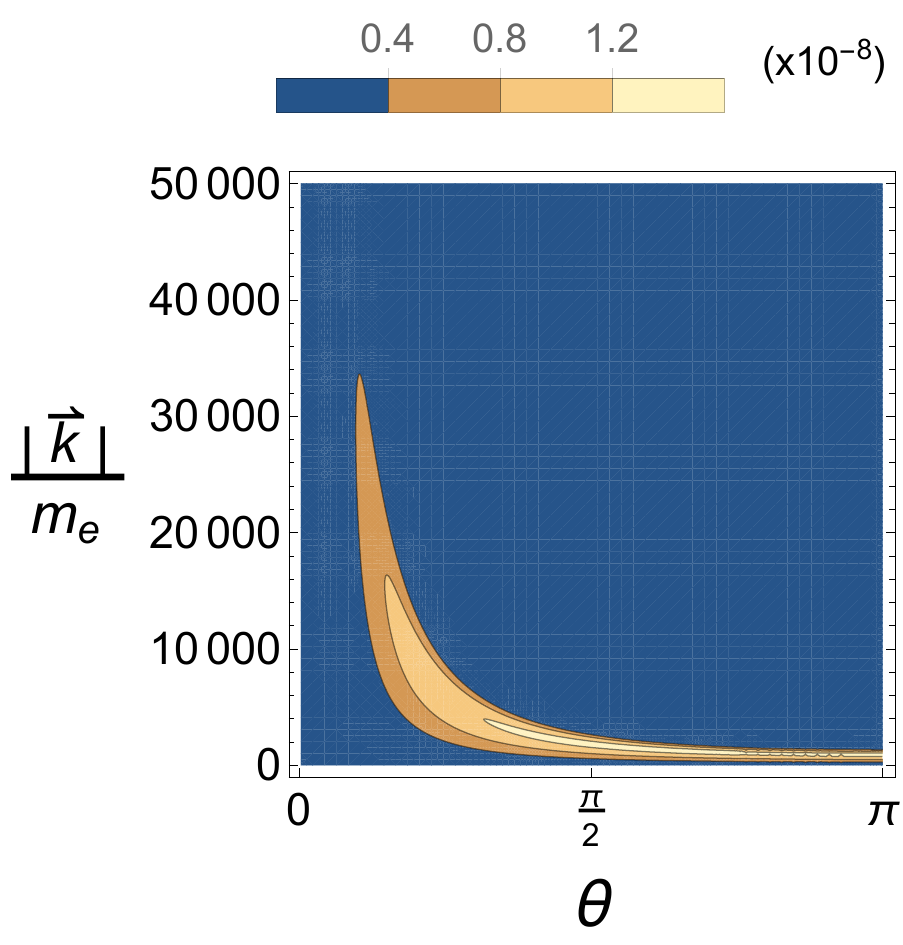}}
}
\subfloat[$P_-,~\delta=0.6$]{
 \raisebox{0.8ex}{
  \includegraphics[scale=0.35]{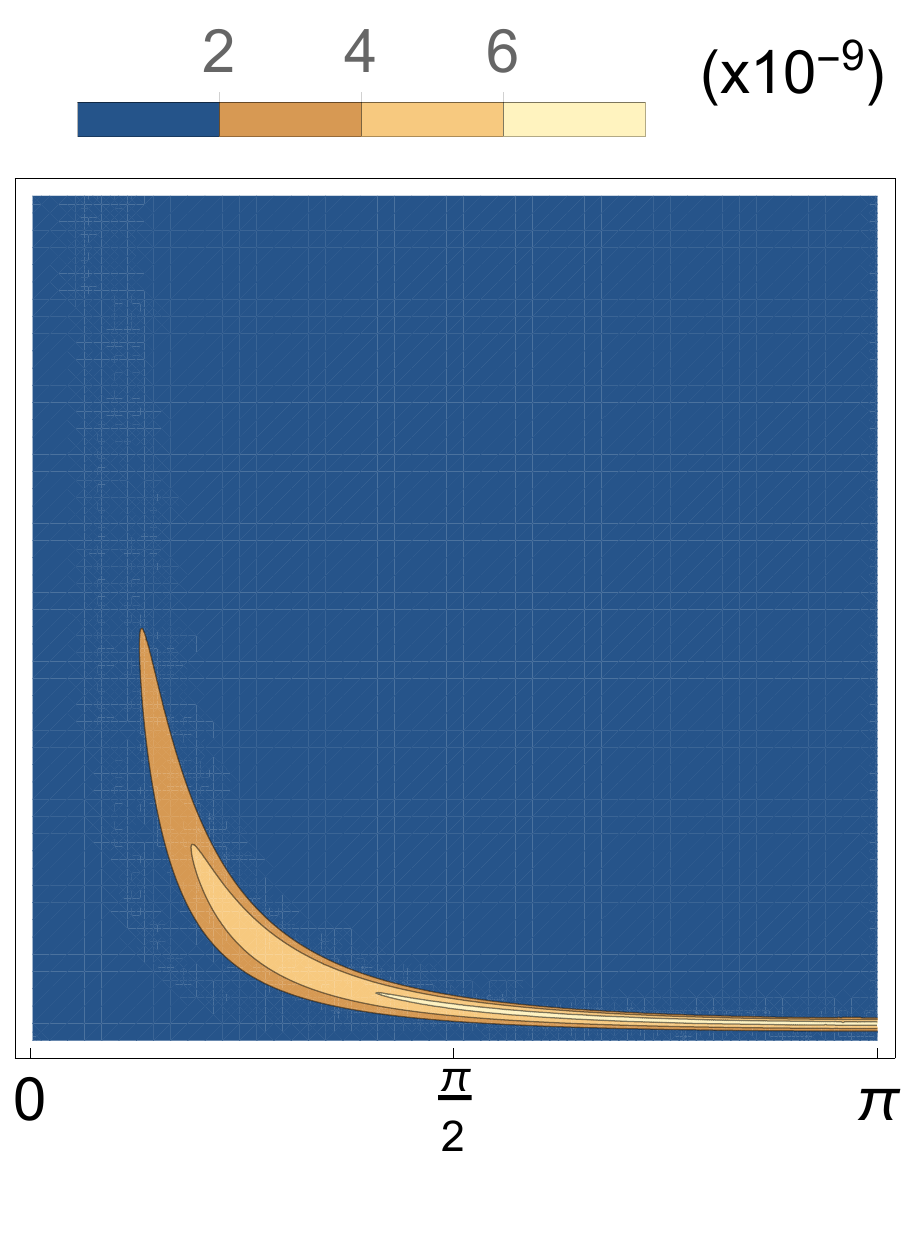}}
}
\subfloat[$P_+,~\delta=0$]{
\raisebox{0.8ex}{
 \includegraphics[scale=0.35]{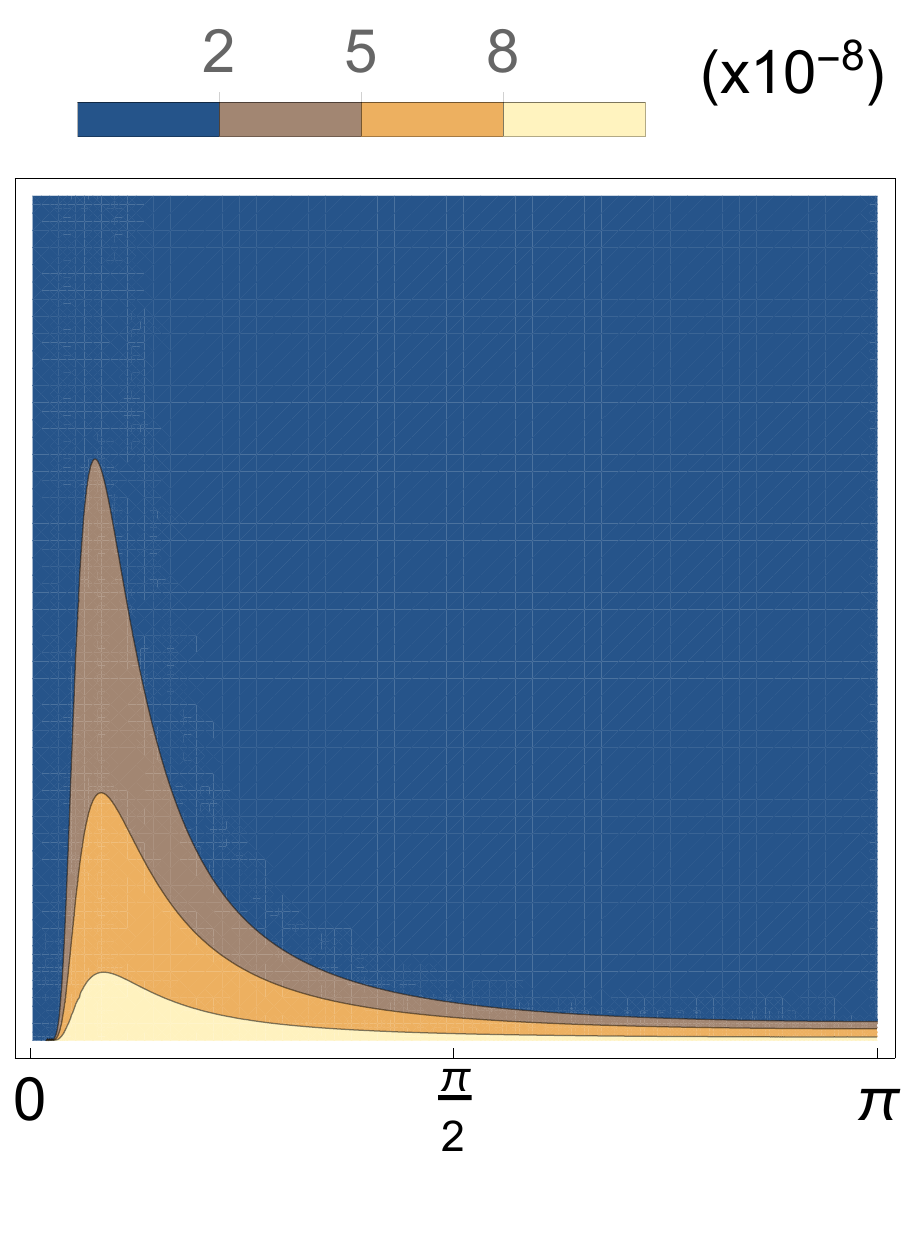}}
}
\subfloat[$P_+,~\delta=0.6$]{
 \raisebox{0.8ex}{
  \includegraphics[scale=0.35]{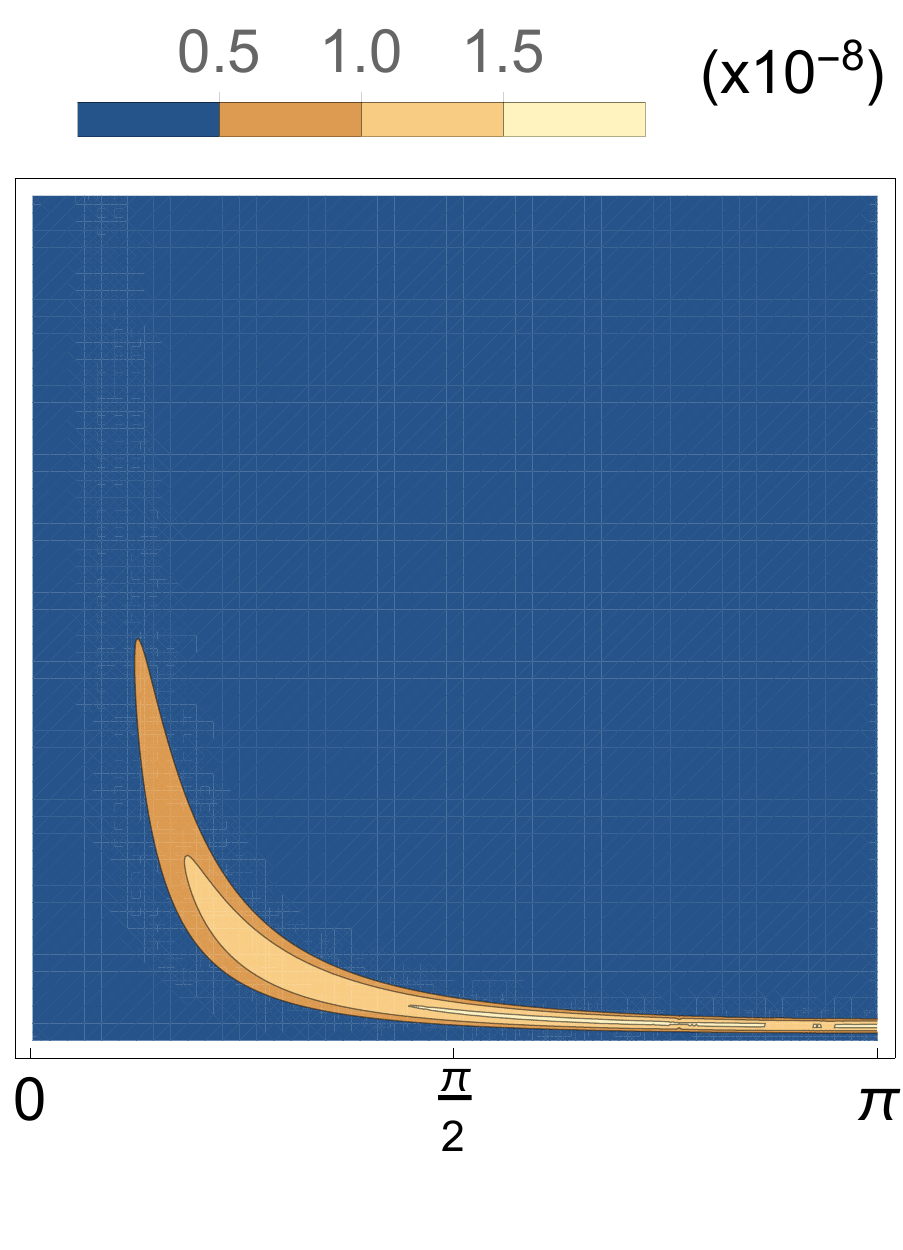}}
}
\caption{The pseudo-scalar (left) and scalar (right) yields, $P_\pm$, are plotted as a function of the polar angle and momenta of the emitted ALP in a Gaussian background for two different ALP masses.  Also, $\chi_{p,0}$ has been fixed to $1$.  \label{kpd}}
\end{figure}

Lastly we show in Figure \ref{kpd} the effects of a non-zero $\delta$ on the angular distribution of emitted ALPs in a Gaussian background with $\chi_{p,0}=1$.

In all of these cases we see that for larger values of the ALP mass the properties of scalar and pseudo-scalar emission become very alike.
This is hinted at by the expressions in Eq. \ref{lcfaP} where we see that the difference between the analytic formulas for pseudo-scalar and scalar emission lies in the $\delta$-dependent pre-factor of the $\text{Ai}_1$ function.

\subsection{ALP production in a high-intensity laser pulse} \label{yieldLaser}

\begin{figure}[H]
\centering
\subfloat[]{
\raisebox{0.1ex}{
 \includegraphics[scale=0.545]{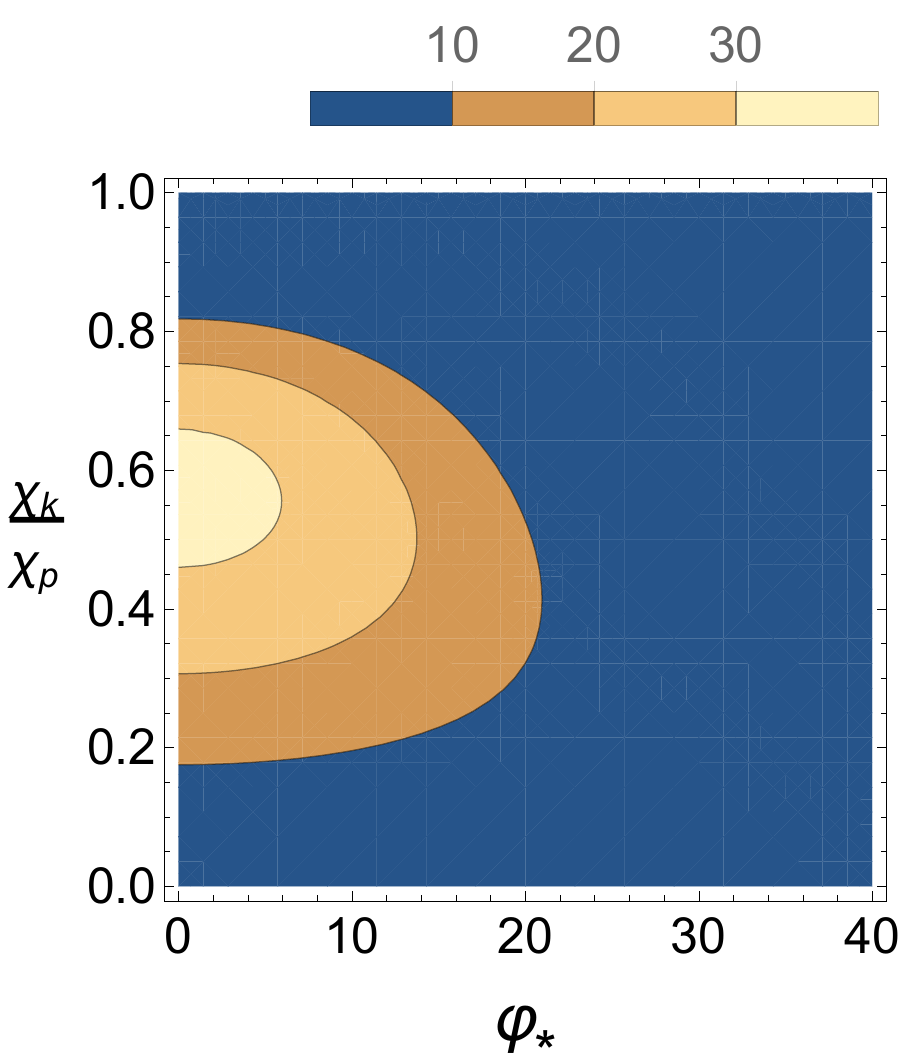}}
}
\subfloat[]{
 \raisebox{0.7ex}{
  \includegraphics[scale=0.453]{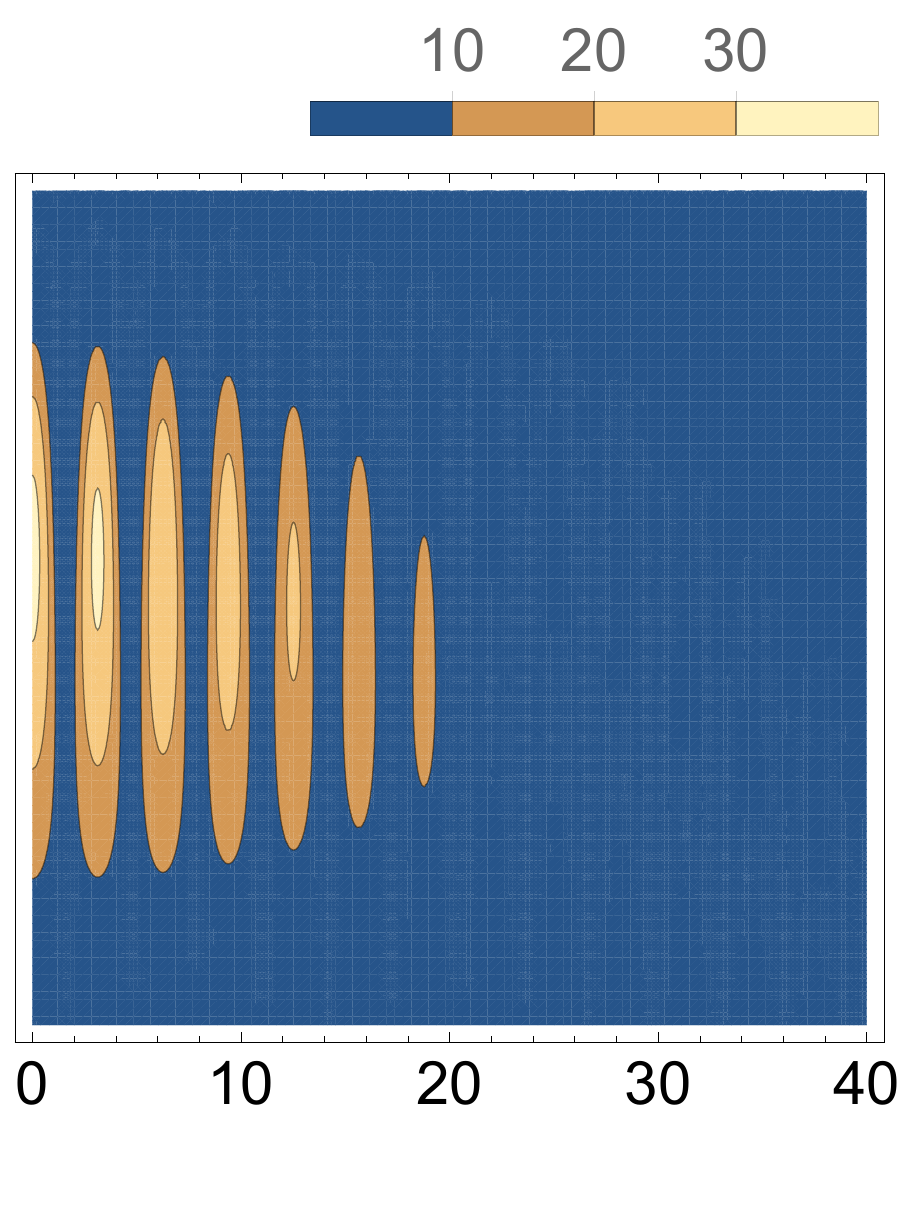}}
}
\caption{The pseudo-scalar yield, $P_-$, is plotted as a function of $\varphi_{\ast}$ and $\chi_{k}/\chi_p$ for $\delta=0$ in (a) a Gaussian pulse background and (b) a laser pulse background, with a pulse duration of $10$ fs and $\kappa^0=1.55$ eV.\label{cpGC1}}
\end{figure}

\noindent The external field backgrounds studied in the previous subsections serve as a useful test case for the physics of ALP production in more general external fields.
However in a laser-based experimental set-up the strong electromagnetic field will have a carrier wave frequency. Assuming that the photons are linearly polarised, an example of this is to take
\beq
g^\prime(\varphi_*)=e^{-\left(\frac{\varphi_*}{\Phi}\right)^2}\cos\varphi_*
\eeq
where we have assumed a Gaussian pulse shape.
In using this pulse shape, we expect the results to be similar to that in the previous section where we simply had a Gaussian profile.
The main difference in a laser pulse background is that the cosine modulation results in a modulation of the differential yield in $\varphi_{\ast}$.
We can see an example of this in Figure \ref{cpGC1} where we plot the $\chi_k$ distribution for pseudo-scalar emission with $\chi_{p,0}=1$ and $m_{\phi}=0$.  For Figure \ref{cpGC1} we choose a shorter pulse duration of $10$ fs such that the modulation effects are more apparent, as for longer pulses the wavelength in units of $\varphi_*$ becomes very small in comparison the the pulse duration in units of $\varphi_*$.

We can see that in the case of the laser pulse background the overall shape of the contours matches that in the case of just a Gaussian background.
It is also important to note that the peak values in the differential yield are also the same.
This tells us that the total yield in both cases will be similar, and in fact the total yield in the case of a laser pulse only differs from that in the Gaussian case by a numerical factor $\sim 50\%$.
The effects of a non-zero mass and varying seed electron momentum mirrors those in Figure \ref{TYmassD} and \ref{TYd0log} for the Gaussian case.

\section{Conclusions, analysis, and outlook}

\noindent In the interest of future lab-based experimental ALP searches a detailed study of ALP production via Compton scattering in low- and high-intensity electromagnetic fields, with a particular focus on laser pulses, was performed.
Particular properties of the production yields depend strongly on the CP nature of the ALP, through which its coupling to electrons is determined.
The basic set-up that we envisage is an electron colliding almost head-on with a laser pulse. For optical lasers, photons have energies of the order of eV, and in an optical set-up, the electron momentum could be anywhere from keV to several GeV.

In Section \ref{perturbative} the production yields for ALPs were derived in the case of a low-intensity laser pulse interacting with an electron.
That is, larger ALP masses suppress the production yield, with the cutoff on the ALP mass imposed by having a non-zero production yield being largely determined by the energy of the photons in the background field.
The angular distributions in Section \ref{DistPert} show that the ALPs with larger energies are emitted in the direction in which the incoming seed electron propagates, whereas lower energy ALPs are emitted off-axes.
We can estimate the number of axions emitted in an interaction between a low-intensity laser pulse and an electron bunch as
\beq \label{lowIntN}
N_{\phi}^{\xi\ll1}=10^{11}Wg_{\phi e}^2\xi^2\left(\frac{N_e}{10^{8}}\right)\left(\frac{\Phi^\prime}{10^3}\right)
\eeq
where $W$ parametrises the kinematical and axion mass dependencies, $N_e\times10^{8}$ is the number of electrons per bunch, and $\Phi=\Phi^\prime\times10^3$.  Typical expectations from a lab-based experimental set-up are that $N_e$ and $\Phi^\prime$ would be $\mathcal{O}(1)$ numbers.
For scalar production we find that $W\simeq0.066$ for light ALPs regardless of what the electron momentum is, however this number begins to decrease for larger ALP masses.
To maintain $W\simeq0.066$ for larger axion masses, up to $\sim0.1m_e$, it is only required to increase the momentum of the electrons interacting with the pulse.  However, for $m_{\phi}>m_e$, $W$ becomes suppressed.  For pseudo-scalars the situation is much different.  For light ALP masses $W$ strongly depends on the momenta of the incoming electrons, where for electrons at rest $W\sim10^{-12}$.  For ultra-relativistic electrons however ($|\vec{p}|\sim10^4m_e$) the $W$ factor for the pseudo-scalars becomes similar to that for the scalars.  The behaviour of $W$ for the ALPs can be determined from Figure~\ref{TYmassPert}.

In Section \ref{strong} the production yields for ALPs in a high-intensity constant-crossed electromagnetic field were derived.
In Section \ref{lcfa} the LCFA was used to translate this result to an approximation for the production yield in non-trivial field configurations, such as a laser pulse.
The sensitivity of the yield to the CP nature of the ALP and the mass of the ALP was studied in detail in Sections \ref{yieldCF}-\ref{yieldLaser}.
It was found that the CP-even states naturally have a larger production yield than the CP-odd states, however as the axion mass is increased the total production yields become qualitatively similar both in magnitude and in their sensitivity to the seed electron energy.
An increase in the ALP mass always leads to a reduction in the production yield, however this is not as drastic as in the low-intensity regime.
Here we find that the production yield remains sizeable even for ALPs with masses greater than that of the electron.
The reason for this lies in the fact that the electron can absorb many photons before emitting an ALP, thus increasing the available energy for ALP production.
Also, regardless of the ALP mass both production yields become similar in magnitude for larger seed electron energies. 
This is quite similar to the behaviour seen in the low-intensity case.
The most interesting effect comes from the angular and momentum distributions of the emitted ALPs -  using a Gaussian pulse for the background field, it was demonstrated that the angular distributions were strongly dependent on the CP nature of the ALP.
We found that CP-odd ALPs have a momentum distribution peaked at some non-zero value determined by the seed electron energy, and the CP-even ALPs have a momentum distribution peaked at zero.
However, allowing the ALP mass to increase to $\mathcal{O}(m_e)$ we find that the angular and momentum distributions for the CP-even and -odd ALPs become virtually indistinguishable, differing only in magnitude.
Due to the non-trivial dependence of the production yield on $\xi$ and $\Phi$ in a high-intensity laser background we cannot factorise these as we have in Eq~\ref{lowIntN} for the low intensity laser pulse.
However, as can be seen from the previous section the production yields are typically at least an order of magnitude larger than in the case of a low-intensity laser pulse, with a larger range of accessible ALP masses.

These results constitute the first analysis of spin-0 particle production via non-linear Compton scattering in intense laser pulses.
In this paper we have not only described how theoretical predictions for such a process are calculated, but we have already obtained information on the characteristics of the production mechanism which will be crucial to understanding how to detect these ALPs in a lab-based experimental set-up.

\section*{Acknowledgments}
\noindent The authors acknowledge funding from Grant No. EP/P005217/1.

\appendix

\section{Light-front coordinates}  \label{appA}
\noindent In light-front coordinates we define
\beq
x^-=x^0-x^3,~~x^+=x^0+x^3,~~x^\perp=(x^1,x^2)
\eeq
and
\beq
x_-=\tfrac{1}{2}x^+,~~x_+=\tfrac{1}{2}x^-,~~x_{\perp}=-x^{\perp}.
\eeq
In this way we have
\beq
x\cd z=x^+z_++x^-z_-+x^\perp\cd z_\perp=\tfrac{1}{2}x^+z^-+\tfrac{1}{2}x^-z^+-x^\perp\cd z^\perp.
\eeq
In the spatial integrals we have
\beq
dx^0dx^3=\tfrac{1}{2}dx^+dx^-
\eeq 
whereas in the momentum integrals we have
\beq
\int \frac{d^3p}{(2\pi)^3}\frac{\theta(p^0)}{p^0}=\int\frac{d^2p^\perp dp^-}{(2\pi)^3}\frac{\theta(p^-)}{p^-}
\eeq
where the on-shell condition changes from
\beq
p^2=m^2~~\rightarrow~~ p^+=\frac{p^\perp \cdot p^\perp+m^2}{p^-}.
\eeq

\subsection{Angular spectra in light-front coordinates} \label{angular}

\noindent We can obtain information on the angular spectra of the emitted scalar or pseudo-scalar in the lab frame by transforming to polar coordinates defined by $k^1=|\vec{k}|\sin\theta\cos\eta$, $k^2=|\vec{k}|\sin\theta\sin\eta$, and $k^3=|\vec{k}|\cos\theta$.
The integration measure then becomes
\beq
\int dk^1dk^2dk^-=-\int d\theta d\eta d|\vec{k}|~|\vec{k}|^2\sin\theta\left(1-\frac{|\vec{k}|}{\sqrt{|\vec{k}|^2+m_\phi^2}}\cos\theta\right).
\eeq
If we are in the situation where we have already integrated out the $k^2$ variable then the relevant coordinate transformation is $k^1=|\vec{k}|\sin\theta$ and $k^3=|\vec{k}|\cos\theta$, from which the integration measure then shifts to
\beq \label{polar}
\int dk^1dk^-=\int d\theta d|\vec{k}|~|\vec{k}|\left(1-\frac{|\vec{k}|}{\sqrt{|\vec{k}|^2+m_\phi^2}}\cos\theta\right).
\eeq
In lab-frame polar co-ordinates the angle $\theta$ is the angle between the direction of the laser and the direction of the emitted particle.
If we assume that the incoming electrons are counter propagating with the laser then we have
\beq
\chi_p=\frac{\xi\kappa^0}{m_e^2}(E_p+|\vec{p}|)~~~\text{and}~~~\chi_k=\frac{\xi\kappa^0}{m_e^2}(E_k-|\vec{k}|\cos\theta)
\eeq
with $E_q=\sqrt{m_q^2+|\vec{q}|}$.
It is clear in this set-up that larger initial electron energies correspond to larger values of $\chi_p$, however for the emitted particle the relationship between $E_k$ and $\chi_k$ depends on the polar angle $\theta$.
For highly relativistic initial momenta ($|\vec{p}|\gg m_e$) we have that
\beq
\chi_p\simeq2\frac{\xi\kappa^0|\vec{p}|}{m_e^2}\left(1+\frac{m_e^2}{4|\vec{p}|^2}\right).
\eeq
If both the initial electron and the emitted ALP are ultra-relativistic the $\chi_k<\chi_p$ kinematical bound becomes $|\vec{k}|(1-\cos\theta)\lesssim2|\vec{p}|$, and when the emitted ALP is ultra-relativistic then we have $|\vec{k}|(1-\cos\theta)\lesssim p^-$. 
We should pay particular attention to the ultra-relativistic limit for the emitted ALP, since we have a perturbativity bound on the applicability of the LCFA at small $\chi_k$, that is $\frac{\xi^3}{\chi_k}\gg 1$ \cite{DiPiazza1}.
In the ultra-relativistic limit we find that the perturbativity bound is
\beq
\frac{\xi^2}{(1-\cos\theta)}\frac{m_e^2}{\kappa^0|\vec{k}|}\gg1.
\eeq
This is more likely to be violated for ultra-relativistic ALPs which are `back-scattered', i.e. $\cos\theta\simeq-1$.
However if we take $\xi\sim100$ and $\kappa^0\sim1.55$ eV this relation is only violated for sub-eV ALPs.
We will keep this in mind and refer to this appendix when analysing our results.

\section{Airy integral identities}  \label{appB}

\noindent In performing the $S$-matrix and phase space integrals in the constant high intensity background we found the following Airy function relations useful \cite{KingTrident}:
\begin{align}
\int_{-\infty}^{\infty}dx~e^{i(cx+dx^3)}=&\frac{2\pi}{(3d)^{1/3}}\text{Ai}\left(\frac{c}{(3d)^{1/3}}\right)	\non\\
\int_{-\infty}^{\infty}dx~x^2e^{i(cx+dx^3)}=&-\frac{2\pi}{(3d)^{1/3}}\frac{c}{3d}\text{Ai}\left(\frac{c}{(3d)^{1/3}}\right)	\non\\
\int_{-\infty}^{\infty}dx~\text{Ai}\left(c+dx^2\right)=&\frac{2^{2/3}\pi}{\sqrt{d}}\text{Ai}^2\left(\frac{c}{2^{2/3}}\right)	\non\\
\int_{-\infty}^{\infty}dx~x\text{Ai}\left(c+dx^2\right)=&-\frac{2^{2/3}\pi}{\sqrt{d}}\frac{c}{2d}\text{Ai}^2\left(\frac{c}{2^{2/3}}\right)	\non\\
\int_{-\infty}^{\infty}dx~x^2\text{Ai}\left(c+dx^2\right)=&-\frac{2^{2/3}\pi}{d^{3/2}}\left[\frac{c}{2}\text{Ai}^2\left(\frac{c}{2^{2/3}}\right)-2^{-1/3}\text{Ai}^{\prime2}\left(\frac{c}{2^{2/3}}\right)\right]	\non\\
\int_{-\infty}^{\infty}dx~(b+dx^2)\text{Ai}^2\left(c+dx^2\right)=&-\frac{1}{4\sqrt{d}}\left[(c-2b)\text{Ai}_1\left(2^{2/3}c\right)+\frac{1}{2^{2/3}}\text{Ai}^{\prime}\left(2^{2/3}c\right)\right]	\non\\
\int_{-\infty}^{\infty}dx~\text{Ai}^{\prime2}\left(c+dx^2\right)=&-\frac{2^{-2/3}}{4\sqrt{d}}\left[3\text{Ai}^{\prime}\left(2^{2/3}c\right)+2^{2/3}c\text{Ai}_1\left(2^{2/3}c\right)\right].
\end{align}

\end{document}